\documentclass[12pt,preprint]{aastex}

\shorttitle{Crystallization by Planetesimal Bow Shocks}
\shortauthors{Miura et al.}

\begin{document}


\title{Formation of Cosmic Crystals
 in Highly-Supersaturated Silicate Vapor 
 Produced by Planetesimal Bow Shocks}


\author{H. Miura\altaffilmark{1}, K. K. Tanaka\altaffilmark{2}, T. Yamamoto\altaffilmark{2}, T. Nakamoto\altaffilmark{3}, J. Yamada\altaffilmark{1}, K. Tsukamoto\altaffilmark{1}, and J. Nozawa\altaffilmark{1,4}}

\altaffiltext{1}{Department of Earth and Planetary Materials Science, Graduate School of Science, Tohoku University, Aoba 6-3, Aramaki, Aoba-ku, Sendai 980-8578, Japan}

\altaffiltext{2}{Institute of Low Temperature Science, Hokkaido University, Sapporo 060-0819, Japan}

\altaffiltext{3}{Earth and Planetary Sciences, Tokyo Institute of Technology, Meguro, Tokyo 152-8551, Japan}

\altaffiltext{4}{Institute for Materials Research, Tohoku University, 2-1-1 Katahira, Aoba-ku, Sendai 980-8577, Japan}

\email{miurah@m.tains.tohoku.ac.jp}


%\altaffiltext{1}{Visiting Astronomer, Cerro Tololo Inter-American Observatory.
%CTIO is operated by AURA, Inc.\ under contract to the National Science
%\altaffiltext{2}{Society of Fellows, Harvard University.}
%\altaffiltext{3}{present address: Center for Astrophysics,
%    60 Garden Street, Cambridge, MA 02138}
%\altaffiltext{4}{Visiting Programmer, Space Telescope Science Institute}
%\altaffiltext{5}{Patron, Alonso's Bar and Grill}


\begin{abstract}
Several lines of evidence suggest that fine silicate crystals observed
in primitive meteorite and interplanetary dust particles (IDPs)
nucleated in a supersaturated silicate vapor followed by crystalline
growth. We investigated evaporation of $\mu$m-sized silicate particles
heated by a bow shock produced by a planetesimal orbiting in the gas in
the early solar nebula and condensation of crystalline silicate from the
vapor thus produced.  Our numerical simulation of shock-wave heating
showed that these $\mu$m-sized particles evaporated almost completely
when the bow shock is strong enough to cause melting of chondrule precursor dust particles.  We found that the silicate vapor cools very rapidly with expansion into the
ambient unshocked nebular region; the cooling rate is estimated, for
instance, to be as high as $ 2000 \ {\rm K \ s^{-1}}$ for a vapor heated
by a bow shock associated with a planetesimal of radius $1 \ {\rm
km}$. The rapid cooling of the vapor leads to nonequilibrium gas-phase
condensation of dust at temperatures much lower than those expected from
the equilibrium condensation.  It was found that the condensation
temperatures are lower by a few hundred K or more than the
equilibrium temperatures. This explains the results of the recent
experimental studies of condensation from a silicate vapor that
condensation in such large supercooling reproduces morphologies similar
to those of silicate crystals found in meteorites. Our results suggest
strongly that the planetesimal bow shock is one of the plausible sites
for formation of not only chondrules but also other cosmic crystals in
the early solar system.
\end{abstract}





\section{Introduction}
\label{sec:introduction}

Vapor-solid (VS) growth is a major process for dust formation in the
inner region of the early solar nebula, where the gas pressure was too
low for a liquid phase to exist stably. Actually, there are several
lines of evidence in interplanetary dust particles (IDPs) and primitive
meteorites that the dust condensed directly from the vapor phase as
suggested for example by enstatite whiskers elongated along the $a$-axis
found in IDPs \citep{bradley83} and $\mu$m-sized polyhedral olivine crystals
with various morphologies found in matrix of primitive meteorite \citep{nozawa09}. 
Hereafter, we refer to these fine crystals as cosmic
crystals. It is an important issue to reveal the formation environment
of these cosmic crystals for understanding the early history of the
solar system.

To reproduce the cosmic crystals, evaporation and condensation
experiments have been performed by many authors so far. \citet{mysen88} carried out experiments in the system composed of
Mg$_2$SiO$_4$-SiO$_2$-H$_2$ in the pressure ranging from $10^{-4}$ to
$10^{4} \ {\rm dyn \ cm^{-2}}$ and in the temperature ranging from $1620
\ {\rm K}$ to $1920 \ {\rm K}$ to determine phase relations of the
system. In these experiments, condensation of MgSiO$_3$ and SiO$_2$
phases took place from the Si-rich vapor, which was produced by
incongruent vaporization of enstatite. \citet{tsuchiyama88} analyzed
these condensates with the use of an analytical transmission electron
microscope (ATEM) and a scanning electron microscope (SEM) to compare
the condensates produced in the experiment with enstatite crystals found
in IDPs \citep{bradley83}. They concluded that the characteristic
textures of clinoenstatite found in IDPs were not reproduced in the
evaporation and condensation experiments by \citet{mysen88}.

Recently, \citet{kobatake08} carried out evaporation and
condensation experiments to investigate a relationship between the
growth conditions and morphologies of cosmic crystals condensed from
highly supersaturated vapors. They used a sphere with forsteritic
composition (Mg$_2$SiO$_4$) as an evaporation source and succeeded in
reproducing various morphologies observed in $\mu$m-sized cosmic olivine
crystals.  They found that types of the morphology depend on the
condensation temperature $T_{\rm c}$. Namely, they found bulky-types at
$T_{\rm c} > 1270 \ {\rm K}$, platy-types at $970 < T_{\rm c} < 1270 \
{\rm K}$, and columnar-needle-types at $770 < T_{\rm c} < 1090 \ {\rm
K}$ under the total pressure of $10^{3}$ to $10^{4} \ {\rm dyn \
cm^{-2}}$. Furthermore, \citet{yamada09} carried out the same experiments
except that the evaporation source has enstatitic composition
(MgSiO$_3$) and succeeded in reproducing enstatite whiskers elongated
along the $a$-axis at $700 < T_{\rm c} < 1150 \ {\rm K}$. They also
reproduced enstatite crystals of platy-types at $1150 < T_{\rm c} < 1300
\ {\rm K}$. Crystals produced in their experiments have morphologies very similar to the cosmic
crystals \citep{bradley83, nozawa09}. 
The condensation temperatures of these cosmic crystal analogs produced in the experiments aforementioned are substantially lower than the temperature of $\sim 1400 \ {\rm K}$, at which forsterite and enstatite condense in equilibrium from the solar nebula gas with the total pressure of $10^{3} \ {\rm dyn \ cm^{-2}}$ \citep[e.g.,]{grossman72}.
The reproduction of the morphologies of cosmic crystals in the experiments suggests that the cosmic crystals were formed in highly
supercooled vapors through nucleation and successive crystal growth; the
condensation temperatures of the cosmic crystals are lower than those
expected from the equilibrium theory by a few hundred degrees or
more. To produce such supercooled silicate vapor, one would require
significant evaporation of silicate dust followed by rapid cooling of
its vapor.

As a possible site of formation of the cosmic crystals, we notice a
localized bow shock produced by a planetesimal revolving in a highly
eccentric orbit in a gas of the solar nebula; the shocked region was
originally proposed as a plausible site for chondrule formation \citep{hood98, ciesla04, hood05, hood09}. When chondrule
precursor dust aggregates enter the bow shock, they do not evaporate
significantly because of their large sizes ($\sim$ mm) but melt then
cool and solidify to form chondrules.  However, small dust particles of
$\mu$m in size will evaporate completely behind the shock front if the
shock is strong enough, and produce silicate vapor \citep{miura05}. The vapor cools
rapidly with expansion in the vicinity of the planetesimal in orbit,
resulting in a highly supercooled state.

In this paper, we examine formation of cosmic crystals in the regions
behind planetesimal bow shocks. The point of our discussion is whether the silicate vapor realizes the highly supercooling in which cosmic crystals of various morphologies are produced, or not. We give an overview of our model in
\S\ref{sec:description}. Section~\ref{sec:vapor_density} describes the
process of dust evaporation by shock-wave heating to evaluate the
evaporation fraction of silicate dust particles in a wide range of the
physical parameters. In \S\ref{sec:expansion}, we investigate expansion
of the silicate vapor behind the bow shock and estimate the cooling
rate. Section~\ref{sec:properties_condensate} examines the properties of
the condensates such as their particle sizes and morphologies expected
from our model and compare the results with those obtained by the
evaporation and condensation experiments. We discuss a comprehensive
scenario on the formations of chondrules, cosmic crystals, and other
materials in chondrites in \S\ref{sec:discussion}.

\section{Outline of the Model}
\label{sec:description}

%% In a manner similar to \objectname authors can provide links to dataset
%% hosted at participating data centers via the \dataset{} command.  The
%% second curly bracket argument is printed in the text while the first
%% parentheses argument serves as the valid data set identifier.  Large
%% lists of data set are best provided in a table (see Table 3 for an example).
%% Valid data set identifiers should be obtained from the data center that
%% is currently hosting the data.
%%
%% Note that AASTeX interprets everything between the curly braces in the 
%% macro as regular text, so any special characters, e.g. "#" or "_," must be 
%% preceded by a backslash. Otherwise, you will get a LaTeX error when you 
%% compile your manuscript.  Special characters do not 
%% need to be escaped in the optional, square-bracket argument.

As stated in \S\ref{sec:introduction}, a possible site for formation of
fine cosmic crystals is a localized bow shock region associated with a
planetesimal orbiting in an eccentric orbit.  There arises relative
velocity between a planetesimal and the nebular gas both orbiting around
the Sun, where the eccentricity of the planetesimal orbit is large.
\citet{weidenschilling98} showed that the Jupiter mean motion
resonances can excite planetesimal eccentricities up to $e \simeq 0.3$
or more. \citet{nagasawa05} analyzed the orbital evolution of
terrestrial planetary embryos including the effect of the sweeping
Jupiter secular resonance combined with tidal drag during dissipation of
the protoplanetary gas disk. They found that the eccentricities of
planetary embryos with mass of $0.01 M_{\rm E}$ are excited up to $e
\simeq 0.6$ or more at maximum and oscillate around the mean value of $e \simeq 0.3 - 0.4$ if Jupiter has the eccentricity of $e_{\rm J} = 0.05$ (the
current value is $e_{\rm J} = 0.0485 $), where $M_{\rm E}$ is the Earth
mass. The relative velocity between the eccentric planetesimal and the
circularly orbiting nebular gas is estimated to be $v_{\rm p} \simeq
\sqrt{e^2 + i^2} \ v_{\rm K}$, where $e$ and $i$ are, respectively,
eccentricity and inclination of a planetesimal orbit and $v_{\rm K}$ is
its Keplerian velocity. For $i \ll e$, we obtain
\begin{equation}
v_{\rm p} = 10.3
\left( \frac{ e }{ 0.6 } \right)
\left( \frac{ a }{ 3 \ {\rm AU} } \right)^{-1/2}
\ {\rm km \ s^{-1}},
\label{eq:shock_velocity}
\end{equation}
where $a$ is semi-major axis of a planetesimal orbit. The supersonic velocities relative to the nebula gas produce bow shocks in front of the planetesimals \citep{hood98, ciesla04}. The relative velocity of $v_{\rm p} = 10.3 \ {\rm km \ s^{-1}}$ for $e = 0.6$ at $a = 3 \ {\rm AU}$ is large enough to cause melting of mm-sized silicate dust aggregates \citep{iida01} and evaporation of $\mu$m-sized dust particles \citep{miura05} in a region of the asteroid belt, where the gas density is $n_0 \sim 3\times10^{13} \ {\rm cm^{-3}}$ at the midplane of the gas disk in the minimum mass solar nebula model \citep{hayashi85}.

\begin{figure}
\epsscale{.90}
\plotone{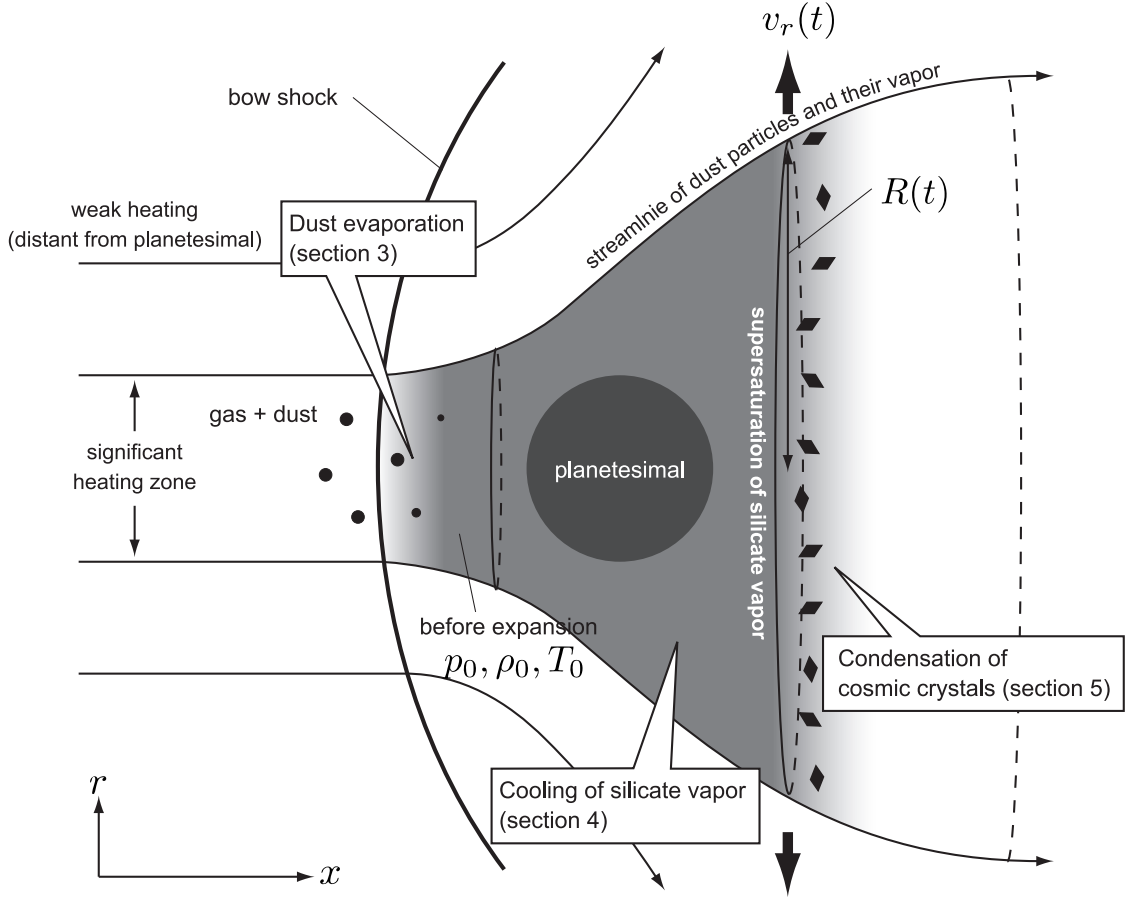}
 \caption{ Outline of the model for cosmic crystal formation. The
 nebular gas and precursor silicate dust come from the left side of the
 planetesimal and pass through the shock front produced by a
 planetesimal orbiting at supersonic velocity in the nebular gas. They
 are heated behind the shock front, and evaporation of the dust
 particles produce silicate vapor, which comes mainly from evaporation
 of $\mu$m-sized particles. The grayed region indicates existence of the vapor. Pressure, density, and temperature of the gas in the
 shocked region just before the expansion ($t = 0$) are denoted by
 $p_0$, $\rho_0$, and $T_0$, respectively.  The shocked region has
 higher pressure than the ambient region and expands vertically with
 velocity $v_{r}$. $R(t)$ is radius of the gas cloud at time $t$. The
 silicate vapor cools with expansion and becomes
 supercooled. Cosmic crystals condense from the cooled vapor after the vapor becomes supersaturated to a certain degree.}
\label{fig:expansion_model}
\end{figure}

An outline of the model for cosmic crystal formation is illustrated in
Fig.~\ref{fig:expansion_model}.  When the shock front formed by
supersonic orbital motion of a planetesimal passes through a region in
the nebula, the nebular gas is abruptly accelerated whereas $\mu$m-sized
dust particles tend to keep their initial position because of their
relatively large inertia. As a result, the dust particles find that they
are exposed to a high-velocity gas flow suddenly and are heated to their
evaporation temperature if the relative velocity and the gas density are
large enough. Evaporation of the $\mu$m-sized dust particles in the
post-shock region will be discussed in \S\ref{sec:vapor_density}.  The
silicate vapor thus produced expands outward because its pressure is
higher than that of the ambient unshocked region. The cooling associated
with the expansion will produce supercooled silicate vapor. We discuss
the cooling process of the silicate vapor and its cooling rate in
\S\ref{sec:expansion}.  The present model supposes that the cosmic
crystals observed in meteorites are condensation products in the
expanding silicate vapor supercooled behind the planetesimal bow
shock. Their sizes, morphologies, and condensation temperatures strongly
depend on the density of silicate vapor and the cooling rate. We shall
show that various kinds of cosmic crystals observed in meteorites are
formed in the cooling of the silicate vapor produced by planetesimal bow
shocks (\S\ref{sec:properties_condensate}).  We shall point out that the
present model leads to simultaneous formation of chondrules and fine
cosmic crystals (\S\ref{sec:discussion}), and that their formation is an
inevitable consequence of formation of planetary systems.

%% In this section, we use  the \subsection command to set off
%% a subsection.  \footnote is used to insert a footnote to the text.


%% This section also includes several of the displayed math environments
%% mentioned in the Author Guide.

\section{Evaporation of $\mu$m-Sized Dust Particles}
\label{sec:vapor_density}

\subsection{Evaporation fraction}
\label{sec:numerical_simulation}

We carry out numerical simulations of shock-wave heating by using a
one-dimensional plane-parallel model developed by \citet{miura06}. Actually, the structure of nebular gas around a planetesimal is
not of one-dimensional plane-parallel structure. However, the
two-dimensional hydrodynamic simulation by \citet{ciesla04} showed
that the one-dimensional plane-parallel approximation was valid in the
vicinity of a planetesimal we are concerned with, say, a few times of
planetesimal radius $R_{\rm p}$. We set a computational domain along the
$x$-axis to be $-R_{\rm p} \le x \le R_{\rm p}$, where the $x$-axis is
parallel to the gas flow and $x = 0$ at the shock front. In this region,
the shocked gas structure can be regarded as one-dimensional
plane-parallel \citep{hood09}.  The simulations were carried out
with varying the following input parameters: the planetesimal radius
$R_{\rm p}$, the pre-shock gas number density $n_0$, the dust-to-gas
mass ratio $\xi$, and the shock velocity $v_{\rm s}$. In the
simulations we set the ranges of parameters to be $1 \le R_{\rm p} \le
1000 \ {\rm km}$, $10^{13} \le n_0 \le 10^{15} \ {\rm cm^{-3}}$, $0.01
\le \xi \le 0.1$, and $5 \le v_{\rm s} \le 60 \ {\rm km}$,
respectively. We take particle radius to be $a_{\rm d} = 1 \ {\rm \mu
m}$ as a typical size of fine dust particles; a scaling to other sizes
is easily done with the use of Eq.~(\ref{eq:evaporation_fraction-text}). 
The case of $\xi = 0.1$ is investigated to see the dependence of the evaporation fraction on $\xi$, although it will require significant settling or concentration of dust particles.

\begin{figure}
\epsscale{.60}
\plotone{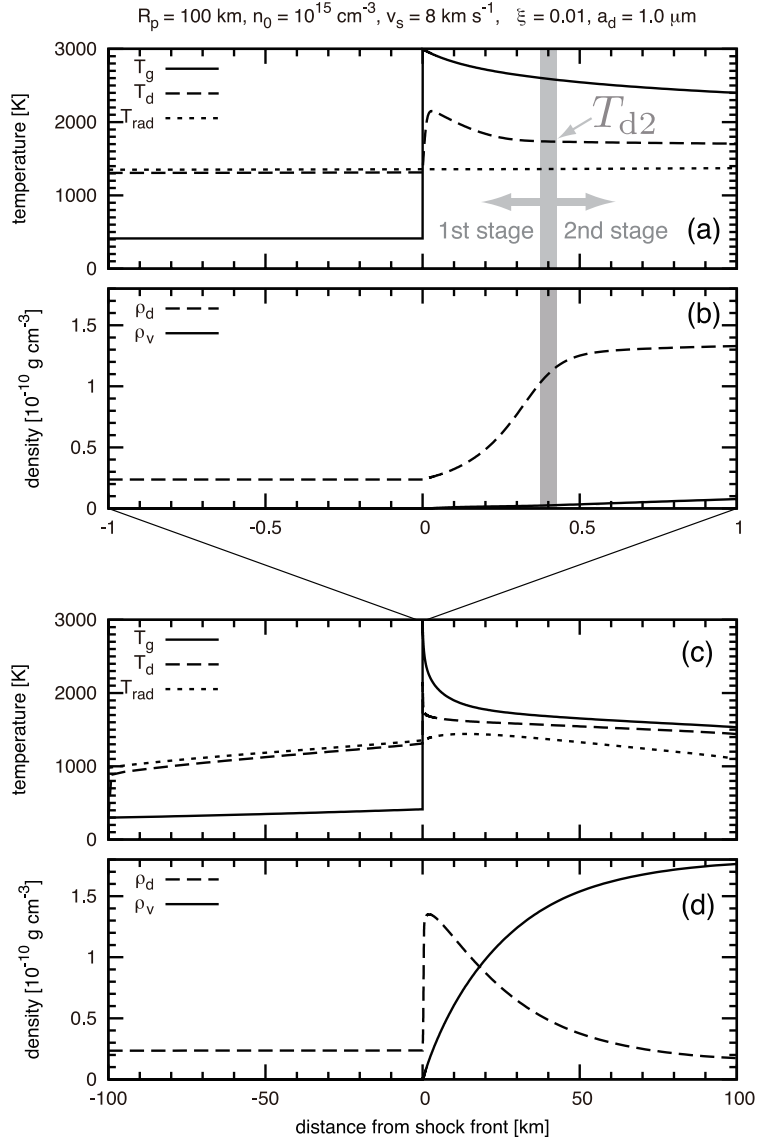}
\caption{%
Spatial profiles of (a)~temperatures ($T_{\rm g}$: gas, $T_{\rm d}$: dust, $T_{\rm rad}$: ambient radiation) and (b)~densities ($\rho_{\rm d}$: dust, $\rho_{\rm v}$: silicate vapor) in the vicinity of the shock front for $R_{\rm p} = 100 \ {\rm km}$, $n_0 = 10^{15} \ {\rm cm^{-3}}$, $v_{\rm s} = 8 \ {\rm km \ s^{-1}}$, $\xi = 0.01$, and $a_{\rm d} = 1 \ {\rm \mu m}$. Panels~(c) and (d) are, respectively, expansions of panels~(a) and (b) in the distance scale. $T_{\rm d2}$ in panel (a) indicates the dust temperature at the beginning of the second stage (see text). 
}
\label{fig:profile_01}
\end{figure}

Figure~\ref{fig:profile_01} shows the result for $R_{\rm p} = 100 \ {\rm
km}$, $n_0 = 10^{15} \ {\rm cm^{-3}}$, $v_{\rm s} = 8 \ {\rm km \
s^{-1}}$, and $\xi = 0.01$. Panel~(a) shows temperature profiles
of the gas ($T_{\rm g}$, solid line), the dust particles ($T_{\rm d}$,
dashed), and the radiation field ($T_{\rm rad}$, dotted) in the
vicinity of the shock front ($-1.0 \le x \le 1.0 \ {\rm km}$). Dust
temperature increases rapidly just behind the shock front by gas
frictional heating due to the velocity difference between gas and dust.
In this stage, which we call the first stage, the dust temperature is
determined by a balance among frictional heating, radiative cooling, and
interaction with the ambient radiation field. The first stage ceases in
a short period of time (less than $0.1 \ {\rm s}$ in this case) because
dust particles come to stop relative to the ambient gas.  Panel~(b)
shows density profiles of dust ($\rho_{\rm d}$, dashed) and silicate
vapor ($\rho_{\rm v}$, solid), which is produced by evaporation of the
dust. It is found that the dust density $\rho_{\rm d}$ increases behind
the shock front ($0 \la x \la 0.3 \ {\rm km}$) because of deceleration
by the gas friction. On the other hand, the vapor density $\rho_{\rm v}$
remains much smaller than $\rho_{\rm d}$, indicating that evaporation of
the dust particles during the first stage is negligible because of the
very short duration of the first stage.

Panel~(c) shows temperature profiles over a wide region around the
planetesimal. The relative velocity between the gas and the dust
particles is almost zero in almost all the region shown here ($x \ga 0.5
\ {\rm km}$), so the frictional heating does not work. However, the dust
temperature is kept above $1500 \ {\rm K}$ because of efficient
collisional heating by the ambient hot gas; we call this stage of
collisional heating the second stage. During the second stage, dust
particles continue to evaporate gradually as is seen from the density
profiles shown in panel~(d). One should note that evaporation of dust
occurs mainly in this stage.  At the edge of the calculation zone ($x =
100 \ {\rm km}$), the vapor density is $\rho_{\rm v} = 1.76 \times
10^{-10} \ {\rm g \ cm^{-3}}$, while the density of survived solid dust
particles is $\rho_{\rm d} = 1.73 \times 10^{-11} \ {\rm g \ cm^{-3}}$.
Therefore, the evaporation fraction $\eta$ defined by
\begin{equation}
\eta
= \frac{ \rho_{\rm v} }{ \rho_{\rm d} + \rho_{\rm v} }
\end{equation}
equals 0.91 in the case shown in Fig.~\ref{fig:profile_01}, implying
that 90\,\% of the dust mass evaporate during the second stage.

\begin{figure}
\epsscale{.80}
\plotone{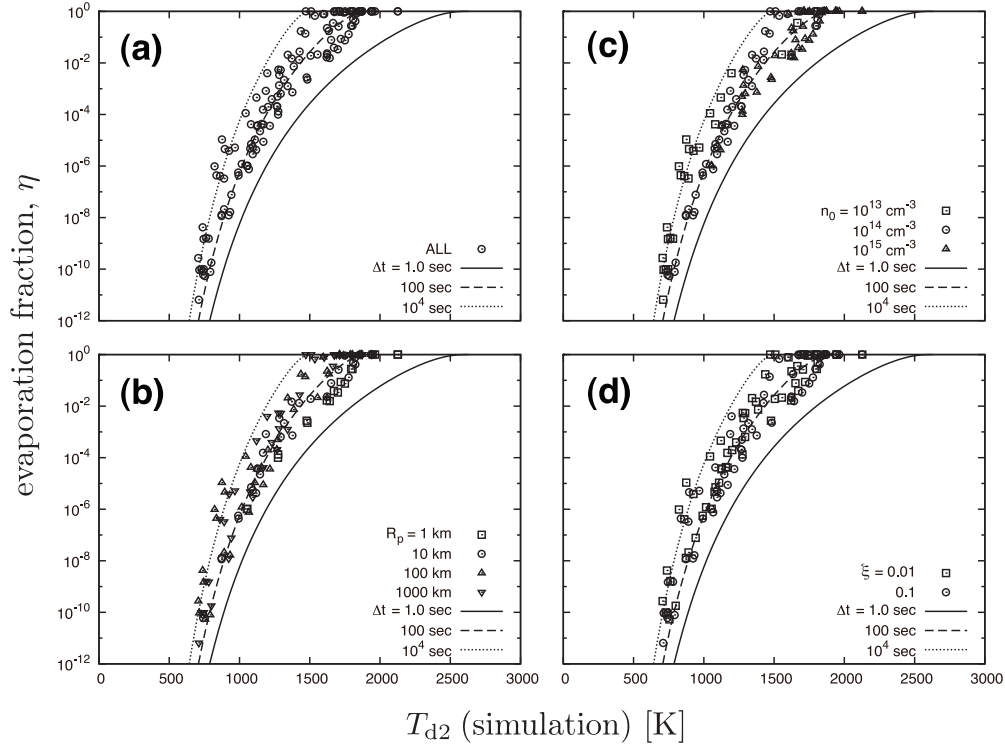}
\caption{
Evaporation fraction $\eta$ as a function of dust temperature at the second stage, $T_{\rm d2}$. We take $T_{\rm d2}$ at the time when the velocity of the dust particles relative to the gas is 1/10 times thermal velocity of the gas molecules.  The symbols indicate numerical results and the curves show $\eta$ calculated with use of Eq.~(\ref{eq:evaporation_fraction-text}) together with Eq.~(\ref{deltaa-fin-text}) for given $\Delta t$, cooling timescale of the gas. Plotted in the panels are (a)~numerical data for all sets of the parameters, (b)~those distinguished by the $R_{\rm p}$-values given in the panel, (c)~those by $n_0$, and (d)~those by $\xi$. 
}
\label{fig:evaporation20090703}
\end{figure}

Figure \ref{fig:evaporation20090703} summarizes the evaporation fraction
$\eta$ for various sets of values of the input parameters. Here, $\eta$
is shown as a function of dust temperature at the second stage. We take
the dust temperature $T_{\rm d2}$ at the time when $v_{\rm rel} / v_T =
0.1$, where $v_T$ is root mean square of thermal velocities of the gas molecules and $v_{\rm rel}$ is the velocity of dust particles relative to the gas; the
result changes little even if we take $v_{\rm rel} / v_T = 0.05$. Open
circles in panel~(a) show numerical results for all sets of the input
parameters.  Figure \ref{fig:evaporation20090703} indicates that $\eta$
increases rapidly with increasing $T_{\rm d2}$.  The temperature
dependence of $\eta$ is given by
\begin{equation}
\eta
= 1 - \left( 1 - \frac{ \Delta a }{ a_{\rm d} } \right)^3 ,
\label{eq:evaporation_fraction-text}
\end{equation}
where $\Delta a$ is size decrease of a dust particle after finishing
substantial evaporation and is given by
\begin{equation}
\Delta a
= \frac{ j_{\rm evap} (T_{\rm d2}) }{ \rho_{\rm c} }
\frac{T_{\rm d2}}{H}
\frac{T_{\rm d2}}{-(dT/dt)_{T_{\rm d2}} } 
\label{deltaa-fin-text}
\end{equation}
for a spherical dust particle (see Appendix
\ref{sec:analytic_formula_evaporation_fraction} for the
derivation). Here,
$\rho_{\rm c}$ is material density of the dust particle, $j_{\rm evap}$
is the evaporation rate, i.e.~mass of vapor evaporating from unit
surface area of the particle per unit time, and $H$
is latent heat of evaporation divided by the gas constant. Note that $\Delta a$ is independent of the original size
$a_{\rm d}$ and $\eta$ is small for a large dust particle. The factor
\begin{equation}
\Delta t_{\rm evap}
= \frac{T_{\rm d2}}{H} \Delta t
\end{equation}
indicates an effective duration of evaporation during the cooling
timescale defined by
\begin{equation}
\Delta t = \frac{T_{\rm d2}}{-(dT/dt)_{T_{\rm d2}}}
\end{equation}
at $T = T_{\rm d2}$. Note that the numerical results are reproduced well
by Eqs.~(\ref{eq:evaporation_fraction-text}) and
(\ref{deltaa-fin-text}).  This implies that the evaporation fraction
$\eta$ mainly determined by dust temperature $T_{\rm d2}$ in the second
stage in spite that there are many other factors ($R_{\rm p}$, $n_0$,
$v_{\rm s}$, and $\xi$) that may affect evaporation of the dust
particles behind planetesimal bow shock.

The duration of substantial evaporation $\Delta t_{\rm evap}$ is
proportional to the cooling timescale $\Delta t$ of the hot gas.  The
expression of $\Delta t$ is very complex in general because it depends
on various physical processes such as vibrational/rotational transitions
of CO and H$_2$O molecules, thermal dissociation of H$_2$ molecules,
Lyman-$\alpha$ emission, and so forth. For a gas of the solar abundance,
however, the major cooling process is Lyman-$\alpha$ emission for $T \ga
10^4 \ {\rm K}$ and thermal dissociation of H$_2$ molecules for $T \ga
3000 \ {\rm K}$ \citep[see Fig.~7]{miura05}.  The timescale
of cooling due to Lyman-$\alpha$ emission is shorter than $\sim 100 \
{\rm s}$.  Below $3000 \ {\rm K}$, the gas cools within a timescale of
$\sim 100 \ {\rm s}$ due to vibrational/rotational transitions of CO and
H$_2$O molecules \citep[see Fig.~7]{miura05}. The cooling
timescale of the hot gas does not depend on the number density of the
gas significantly. In the present case, the cooling timescale of $\Delta
t \sim 100 \ {\rm s}$ reproduces the numerical results well as is seen
from Fig.~\ref{fig:evaporation20090703}.

Finally, let us examine the dependences of the evaporation fraction
$\eta$ on the parameters other than the temperature. Panels~(b), (c),
and (d) examine the dependence of the evaporation fraction $\eta$ on
$R_{\rm p}$, $n_0$, and $\xi$, respectively. In panel~(b), $\eta$
for $R_{\rm p} = 1$, $10$, $100$, and $1000 \ {\rm km}$ are plotted by
different symbols to see the dependence of $\eta$ on $R_{\rm p}$. There
seems no clear systematic dependence of $\eta$ on $R_{\rm p}$ even if we
vary $R_{\rm p}$ by three orders of magnitude. Panel~(c) examines the
dependence on $n_0$, the number density of pre-shock gas.  There seems
to be a slight trend that $\eta$ decreases with increasing $n_0$ but the
dependence is unremarkable compared with the scatter of the data for
each value of $n_0$. Panel~(d) examines the dependence on the
gas-to-dust mass ratio ranging from $\xi = 0.01$ to $0.1$ but we
found no systematic trend of $\eta$ on $\xi$, neither, within the
plausible range of $\xi$.

\subsection{Analytic estimation of the dust temperature}
We have shown that the evaporation fraction $\eta$ is determined mainly
by the dust temperature $T_{\rm d2}$ in the second stage.  However, one
needs to elaborate numerical simulations to calculate $T_{\rm
d2}$. Instead, we derived an approximate analytic expression
(\ref{eq:dust_temp_cond}) of $T_{\rm d2}$ in Appendix
\ref{sec:thermal_conduction} by considering the energy balance of a dust
particle in the second stage. The analytic formula of $T_{\rm d2}$ will
also be useful for calculating the dust temperature and its evaporation
in a planetesimal bow shock in general.

\begin{figure}
\epsscale{.80}
\plotone{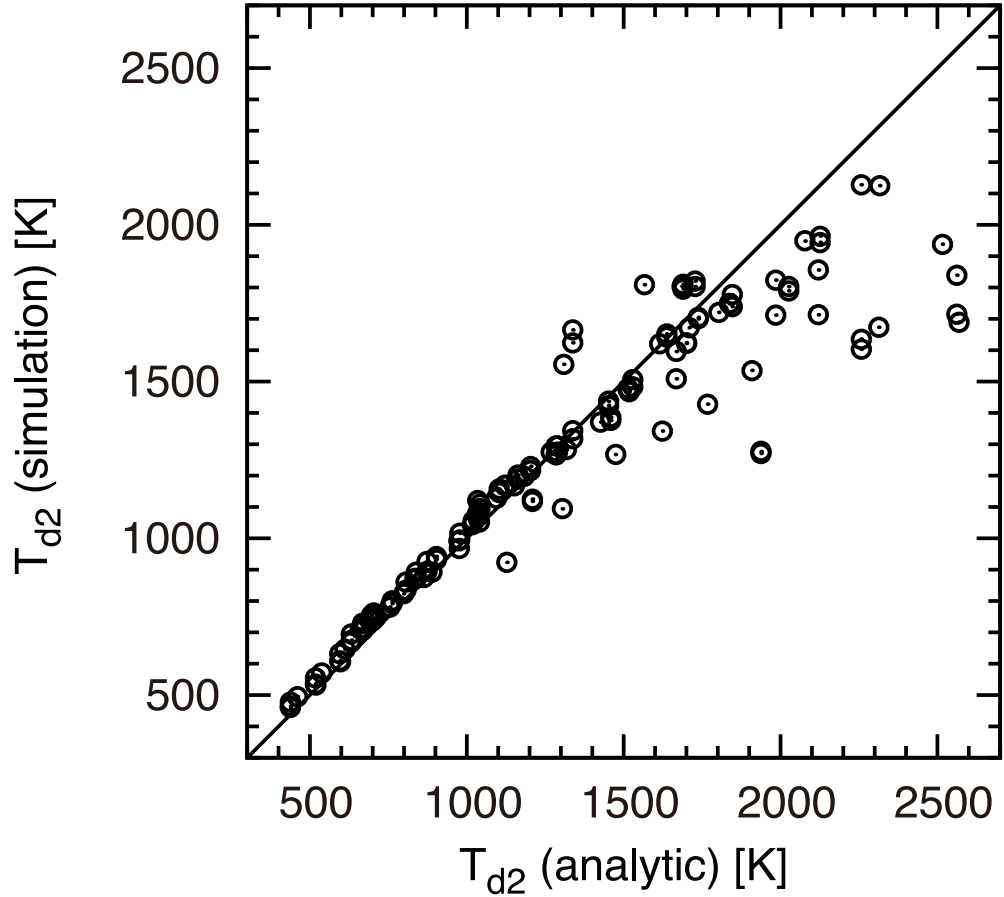}
\caption{
Comparison of $T_{\rm d2}$, dust temperature in the second stage, obtained by the numerical simulations (vertical axis) and those calculated with the use of Eq.~(\ref{eq:dust_temp_cond}) given in Appendix \ref{sec:thermal_conduction} (horizontal axis).
}
\label{fig:Tcond}
\end{figure}

Figure~\ref{fig:Tcond} compares $T_{\rm d2}$ given by
Eq.~(\ref{eq:dust_temp_cond}) with that obtained from the numerical
results. It is found that both agrees with the difference less than $\pm
50 \ {\rm K}$ for $T_{\rm d2} \la 1500 \ {\rm K}$. For $T_{\rm d2} \ga
1500 \ {\rm K}$, the numerical values of $T_{\rm d2}$ are systematically
lower than those given by Eq.~(\ref{eq:dust_temp_cond}). The reason of
the deviation is that the analytic estimation ignores decrease in the
optical depth due to dust evaporation in the shocked region. Actually,
the decrease in the optical depth weakens the intensity of the ambient
radiation field, which heat the dust. In consequence, the dust
temperature decreases and its evaporation is suppressed. This negative
feedback taken into account in the numerical simulation results in the
numerical value of $T_{\rm d2}$ lower than that of the analytic
estimation. The deviation at $T_{\rm d2} \ga 1500 \ {\rm K}$, however,
does not influence the estimation of the evaporation fraction $\eta$
much because $\eta \simeq 1$ in any case at these temperatures as seen
from Fig.~\ref{fig:evaporation20090703}.

\begin{figure}
\epsscale{.80}
\plotone{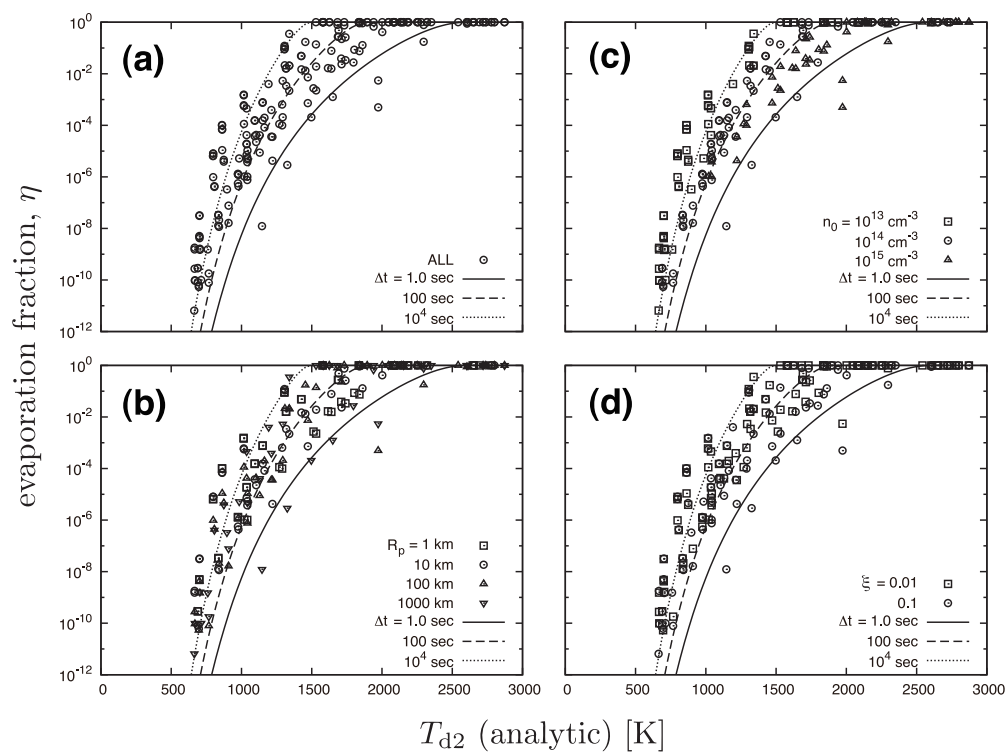}
\caption{
The same as Fig. \ref{fig:evaporation20090703} but the values of $T_{\rm d2}$ in the horizontal axis are replaced by those calculated by using Eq.~(\ref{eq:dust_temp_cond}).
}
\label{fig:evaporation20090701}
\end{figure}

Figure~\ref{fig:evaporation20090701} shows the evaporation fraction
$\eta$ as a function of dust temperature $T_{\rm d2}$ as does
Fig.~\ref{fig:evaporation20090703} but $T_{\rm d2}$ in the horizontal
axis is replaced by the one calculated by using
Eq.~(\ref{eq:dust_temp_cond}). Although the scatter of the data plotted
is larger than in Fig.~\ref{fig:evaporation20090703}, we see that the
analytic formulae still reproduce the evaporation fraction $\eta$.

\section{Expansion and Cooling of the Shocked Gas}
\label{sec:expansion}

\subsection{Equation of expansion}
When the hot gas in the shocked region cools down to the temperatures
lower than $\sim 1500$ K, dust particles re-condense from the vapor
produced by evaporation of the original dust. 
In this subsection, we consider hydrodynamics and cooling of the
expanding gas cloud to characterize the environment for formation of the
cosmic crystals.  Let us assume cylindrical expansion with initial
radius $R_0$ (see Fig.~\ref{fig:expansion_model}). Initial radius of the shocked region $R_0$ is on the same order of magnitude as planetesimal radius $R_{\rm p}$ \citep{ciesla04}. Neglecting the
expansion along the $x$-axis, the expansion velocity $v_{r}$ is
described by
\begin{equation}
\frac{ d v_{r} }{ d t }
= - \frac{1}{\rho} \frac{ \partial p }{ \partial r } ,
\label{eq:eom_r01}
\end{equation}
where $\rho$ is the gas density and $p$ is the gas pressure. We use a
one-zone approximation and approximate $v_{r}$ and $\partial p /
\partial r$ as
\begin{equation}
v_{r} \sim \frac{ dR }{ dt },\quad
- \frac{ \partial p }{ \partial r } \sim \frac{ p }{ R },
\label{eq:one-zone}
\end{equation}
where $R$ is radius of the gas cloud at time $t$. We adopt a polytropic
equation of state for the gas given by
\begin{equation}
p = p_0 \left(\frac{\rho}{\rho_0}\right)^{\gamma},
\label{eq:state}
\end{equation}
where $\rho_0$ is initial gas density and $\gamma > 1$ is a parameter
relating to the polytrope index. The conservation of mass during the
expansion is expressed as
\begin{equation}
R^2 \rho = R_0^2 \rho_0.
\label{eq:density}
\end{equation}
Using Eqs.~(\ref{eq:eom_r01}) to (\ref{eq:density}), we obtain the
equation of expansion of the gas given by
\begin{equation}
\frac{ d^2 \tilde{R} }{ d\tilde{t}^2 }
= \frac{1}{\gamma} \tilde{R}^{-2\gamma+1} ,
\label{eq:eom_r02}
\end{equation}
with $ \tilde{R} = R/R_0$ and $\tilde{t} = (R_0/c_{\rm s0}) t$, where
\begin{equation}
c_{\rm s0} = \sqrt{\frac{\gamma p_0}{\rho_0}}
\label{cs0}
\end{equation}
is sound speed in the gas at $t = 0$. The dimensionless equations for expansion make it clear that the timescale of expansion of the shocked gas behind a planetesimal can be scaled by the sound-crossing time $R_0 / c_{\rm s0}$.

Figure~\ref{fig:expansion} shows the solutions of Eq.~(\ref{eq:eom_r02})
for the initial conditions of $\tilde{R} = 1$ and $\tilde{v}_{r} = 0$
(see Appendix~\ref{sec:analytic_solution}). It is clearly seen that the
expansion is separated into two phases; the acceleration phase, in which
$v_{r}$ increases with time but $R$ remains almost at the initial radius
$R_0$, and the expansion phase, in which the shocked region begins to
expand and $v_{r}$ almost equals a constant terminal velocity. The
dashed curves in panel~(b) show approximations of $v_r$ in the two
limiting cases of $\tilde{t} \ll 1$ and $\tilde{t} \rightarrow \infty$
(see Appendix \ref{sec:analytic_solution}) given by
\begin{equation} 
v_r = \left\{
\begin{array}{ll}
 \displaystyle{\frac{c_{\rm s0}}{\gamma} \frac{t}{t_{\rm s0}}}
& (c_{\rm s0} t \ll R_0), \\
 c_{\rm s0} \displaystyle{\left[\frac{1}{\gamma (\gamma - 1)}\right]^{1/2}}
 & (c_{\rm s0} t \gg R_0), 
\label{vr-limits}
\end{array}
\right.
\end{equation}
where $t_{\rm s0} \equiv R_0/c_{\rm s0}$.

\subsection{Cooling rate of the shocked gas}
\label{sec:cooling_rate}
Using the relation 
\begin{equation}
T
= T_0 (\rho / \rho_0)^{\gamma-1}
= T_0 {\tilde R}^{-2(\gamma - 1)}
\end{equation}
given by Eqs.~(\ref{eq:state}) and (\ref{eq:density}), and $T \propto
p/\rho$, we obtain the time variation of the gas temperature $T$ as
\begin{equation}
-\frac{ dT }{ dt }
= -\frac{dT}{dR} v_{r}
= 2 \left( \frac{\gamma-1}{\gamma} \right)^{1/2}
\frac{T_0 c_{\rm s0}}{R_0} 
{\tilde R}^{-2 \gamma + 1} 
[ 1 - {\tilde R}^{-2(\gamma-1)} ]^{1/2} 
\label{-dT/dt}
\end{equation}
with the use of Eq.~(\ref{eq:vr_solution}) in Appendix
\ref{sec:analytic_solution}.  One sees from Eq.~(\ref{-dT/dt}) that the
cooling rate $-dT/dt$ as a function of ${\tilde R}$ increases with
increasing $R$ at first, reaches a peak, and decreases in proportion to
${\tilde R}^{-2 \gamma + 1}$. 
Figure~\ref{fig:Rcool_temp} shows the cooling rate $-dT/dt$ as a
function of $T$ which decreases monotonically with time.

To evaluate the cooling rate, we need to specify a value of the initial
temperature $T_0$. The gas temperature just behind the shock front could
be higher than $2000 \ {\rm K}$ or more depending on the Mach number
$v_{r}/c_{\rm s0}$. However, even if the temperatures of the gas exceeds
$2000 \ {\rm K}$, it cools rapidly by dissociation of hydrogen molecules
and is kept around $2000 \ {\rm K}$ owing to the energy balance between
re-formation of hydrogen molecules by three-body reaction and their
dissociation \citep{iida01}. We set $T_0 = 2000 \ {\rm K}$ to
estimate the cooling rate around the condensation temperatures.  To
consider condensation through nucleation, on the other hand, we should
refer to the cooling rate $-dT/dt$ when the vapor becomes
supersaturated.  Taking the equilibrium condensation temperatures of
$T_{\rm e} = 1444 \ {\rm K}$ for forsterite and $T_{\rm e} = 1349 \ {\rm
K}$ for enstatite for the total pressure of $10^3 \ {\rm dyn \ cm^{-2}}$
\citep{grossman72} as a measure of estimating the condensation temperature,
we have $T_{\rm e}/T_0 = 0.65 - 0.75$ and $| dT/dt |_{T_{\rm e}} =
(0.25-0.35) T_0/(R_0/c_{\rm s0})$ for $\gamma = 7/5$ and $5/3$ (see
Fig.~\ref{fig:Rcool_temp}). We set $R_0$ to be planetesimal radius
$R_{\rm p}$ in what follows. In consequence, the cooling rate is
estimated to be:
\begin{eqnarray}
-\left(\frac{dT}{dt}\right)_{T_{\rm e}}
&\simeq& \left( 0.25-0.35 \right)
 \frac{ T_0 }{ R_{\rm p} / c_{\rm s0} } \nonumber \\
&\simeq& 2000
\left( \frac{ R_{\rm p} }{ 1 \ {\rm km} } \right)^{-1}
\left( \frac{ T_0 }{ 2000 \ {\rm K} } \right)
\left( \frac{ c_{\rm s0} }{ 3.7 \ {\rm km \ s^{-1}} } \right)
\ {\rm K \ s^{-1}}.
\label{-dT/dt2}
\end{eqnarray}

It should be pointed out that cooling of the shocked gas given by Eq.~(\ref{-dT/dt2}) can be used so far as the pressure of the shocked gas $p$ is much larger than the ambient gas pressure $p_{\rm amb}$. The shocked gas pressure before the expansion is $p \sim 100 \ p_{\rm amb}$ for the shock velocity of an H$_2$ gas of $10 \ {\rm km \ s^{-1}}$. The gas temperature at that time is $\sim 2000 \ {\rm K}$ as a result of the balance between H$_2$ dissociation and its re-formation \citep{iida01, miura05}. The pressure and temperature decrease by subsequent cylindrical expansion. When the temperature drops to the typical condensation temperature of $\sim 1000 \ {\rm K}$, the radius of the cylinder is 2.4 times the initial one for adiabatic expansion, and the shocked gas pressure also decreases to $\sim 1/10$ of that before expansion. However, the gas pressure is still higher than $p_{\rm amb}$ by an order of magnitude. Therefore, Eq.~(\ref{-dT/dt2}) is applicable throughout the expansion phase of interest including the time of condensation.

We focus here the adiabatic expansion because the radiative losses are negligibly small for small shocks as is shown below. Main coolants of the nebula gas at $2000 \ {\rm K}$ are vibrational emissions of CO and H$_2$O molecules. The cooling timescale due to these vibrational emissions was estimated to be $\sim 100 \ {\rm sec}$, which does not significantly depend on the gas density \citep{miura05}. On the other hand, the cooling timescale due to the adiabatic expansion behind a planetesimal is shorter than $\sim 100 \ {\rm sec}$ for planetesimal radius of $ < 100 \ {\rm km}$ (see Eq.~(\ref{-dT/dt2})). Therefore, the shocked gas cools by the expansion before the vibrational emissions remove the thermal energy significantly. The radiative losses might work for large shocks ($\ga 100 \ {\rm km}$) because the expansion takes longer time. However, a large optical depth for these emissions resulting from large shocks prevents the radiative losses from being efficient.

\subsection{Possibility of chondrule formation}
In the formation of chondrules, their cooling rate during solidification is one of the key physical quantities. According to the planetesimal bow shock model, the cooling rate was estimated to be $\sim 10^3 \ {\rm K \ hr^{-1}}$ for planetesimal radius $R_{\rm p} = 1000 \ {\rm km}$, $> 10^4 \ {\rm K \ hr^{-1}}$ for $R_{\rm p} = 100 \ {\rm km}$, and $> 10^5 \ {\rm K \ hr^{-1}}$ for $R_{\rm p} = 10 \ {\rm km}$ \citep{hood05}. On the other hand, the cooling rate of the shocked gas calculated from Eq.~(\ref{-dT/dt2}) is $7\times10^3$, $7\times10^4$, and $7\times10^5 \ {\rm K \ hr^{-1}}$ for $R_{\rm p} = 1000$, $100$, and $10 \ {\rm km}$, respectively. Although Eq.~(\ref{-dT/dt2}) is not a cooling rate of a chondrule itself but of the shocked gas strictly speaking, we note that both estimations of the cooling rates are comparable; this is because the cooling of chondrules is regulated by that of the shocked gas \citep{iida01}. Therefore, Eq.~(\ref{-dT/dt2}) measures the cooling rate of chondrules.

A widely accepted range of the cooling rate of chondrules at solidification is $\sim 10 - 1000 \ {\rm K \ hr^{-1}}$ \citep[and references therein]{hewins05}, which is much slower than that predicted by Eq.~(\ref{-dT/dt2}). However, we consider that this disagreement does not necessarily exclude planetesimal bow shocks as a chondrule formation site. In fact, some authors assert rapid cooling rates, which are in the range estimated from Eq.~(\ref{-dT/dt2}). 
\citet{yurimoto02} proposed that rapid cooling ($\sim 10^5 - 10^6 \ {\rm K \ hr^{-1}}$) was necessary to account for the observed Fe-Mg and O-isotopic exchange in a CO-chondrite type-II chondrule. \citet{wasson03} proposed that very thin overgrowths on some relict grains in chondrules must have been formed by the rapid cooling. The crystallization experiments of a melt droplet by a levitation method succeeded in reproducing chondrule-solidification textures in the experimental conditions of the rapid cooling \citep{tsukamoto99, nagashima06}. Although the rapid cooling scenario does not seem to have been accepted widely to the meteoritic community \citep{hewins05}, there has been no definite evidence that rejects the rapid cooling scenario completely. Therefore, we consider that the planetesimal bow shock is still one of the possible models to be studied for chondrule formation.

\section{Formation of Cosmic Crystals}
\label{sec:properties_condensate}

\subsection{Cooling parameter $\Lambda$ for homogeneous nucleation}
\label{sec:cooling_parameter}
Cosmic crystals condense in the course of cooling of the vapor produced by a planetesimal bow shock. When almost all dust particles evaporate by the bow shock, there is no solid surface available on which the supersaturated vapor condenses. In this case, cosmic crystals are formed through homogeneous (spontaneous) nucleation. In homogeneous nucleation, condensation does not begin when the cooling vapor becomes saturated but begins effectively after the vapor becomes supersaturated to a certain degree.

\citet{yamamoto77} and \citet{draine77} formulated a
grain formation process though homogeneous nucleation in a vapor and
derived analytical expressions of a typical size of grains and their
{\em actual} condensation temperature $T_{\rm c}$ in a supercooling
state as functions of two dimensionless parameters. One is a cooling
parameter defined by
\begin{equation}
\Lambda = \frac{\nu_{\rm coll} t_{T} }{H/T_{\rm e} - 1} ,
\label{eq:lambda01}
\end{equation}
where $\nu_{\rm coll}$ is collision frequency of vapor molecules in
thermal motion, $t_T = T_{\rm e} / \left( -dT/dt \right)_{T_{\rm e}}$ is
cooling timescale of a vapor at $T = T_{\rm e}$ with $T_{\rm e}$ being {\it equilibrium} condensation temperature, and 
$H$ is latent heat of condensation divided by the gas constant and equivalent to that of evaporation (see Appendix \ref{sec:analytic_formula_evaporation_fraction}).
Note that $T_{\rm e}\, (> T_{\rm c})$ is a temperature at
which a vapor and a {\em bulk} condensate co-exit in chemical
equilibrium and approaches $T_{\rm c}$ as $t_T$ gets so long that the
equilibrium between the vapor and the condensate is realized. Grain size $a_{\rm d}$ is mainly determined by $\Lambda$ and is roughly given by $a_{\rm d} / a_0 \sim 0.1 \, \Lambda$ for $\Lambda \gg 1$, where $a_0$ is the radius of a vapor molecule \citep{yamamoto77}. 
In Eq.~(\ref{eq:lambda01}), $\nu_{\rm coll}$ is calculated from the vapor density $\rho_{\rm v}$, and $t_T$ from the cooling rate of the vapor (see Appendix \ref{sec:multi_component}). 
The other parameter is a dimensionless surface tension defined by
\begin{equation}
\Gamma = \frac{4 \pi a_0^2 \gamma_{\rm s}}{k_{\rm B} T_{\rm e}},
\end{equation}
where $\gamma_{\rm s}$ is surface tension of a condensate and $a_0 = (3
\mu_{\rm c} m_{\rm a}/4\pi \rho_{\rm c})^{1/3}$ (i.e.~equivalent radius
of a sphere whose volume equals the volume of a unit cell of the
condensate) with $\mu_{\rm c}$ being molecular weight of a unit cell of
a condensate, $m_{\rm a} = 1.66\times10^{-24} \ {\rm g}$ is atomic mass unit, and $\rho_{\rm c}$ its bulk density. A degree of supercooling
$\Delta T = T_{\rm e} - T_{\rm c}$ is mainly determined by the parameter
$\Gamma$ and is approximately related to $\Gamma$ as $\Delta T \propto
\Gamma^{3/2}$ \citep{yamamoto77}.

Figure~\ref{fig:Lambda} shows a relation between $\Lambda$ and the
evaporation fraction $\eta$. Each of the plots indicates $\eta$
calculated in \S\ref{sec:vapor_density} for a given set of values of the
parameters, while $\Lambda$ is calculated from
Eq.~(\ref{eq:lambda}). All panels indicate the trend that $\Lambda$
increases with $\eta$. This is simply because the larger degree of
evaporation of pre-existing dust is, the larger amount of the vapor is
produced, which in consequence provides favorable conditions for
homogeneous condensation of cosmic crystals. 
Note that homogeneous condensation is possible only if $\Lambda > 1$; otherwise, the vapor is too tenuous for condensation to occur. 
Panel~(a) shows the results of the
calculations for all of the parameter sets, indicating that there appear
many cases of $\Lambda > 1$ for $\eta > 10^{-4}$. Even the cases of
$\Lambda$ as large as $10^5$ are realized for complete evaporation
($\eta \simeq 1$) of pre-existing dust. The contribution to the vapor
production comes mainly from $\mu$m-sized dust particles if their size
distribution is steeper than $a_{\rm d}^{-2}$. The presence of many
cases of $\Lambda > 1$ implies that condensation of cosmic crystals
through homogeneous nucleation behind planetesimal bow shocks is
possible for $\eta > 10^{-4}$. We note that a variety in the
$\Lambda$-values suggests formation of various kinds of cosmic crystals.
The panel~(b) displays the dependence of $\eta$ and $\Lambda$ on the
planetesimal radius $R_{\rm p}$. From panel~(b), one sees that the
homogeneous condensation occurs hardly except for $\eta \sim 1$ for a bow shock
produced by small planetesimals of $R_{\rm p} = 1 \ {\rm km}$ but occurs
almost always for a planetesimal of $R_{\rm p} = 1000 \ {\rm km}$ even if the
evaporation is not so significant ($\eta \ga 10^{-4}$).

\subsection{Size and morphology of cosmic crystals}
\label{sec:averagedsize_condensationtemperature}
%

%Sizes of dust particles condensed through homogeneous nucleation are mainly determined by the cooling parameter $\Lambda$ as stated in \S\ref{sec:cooling_parameter}, while their morphologies depend mainly on supercooling $\Delta T \propto \Gamma^{3/2}$ at which they condense \citep{kobatake08, yamada09}.  

Figure~\ref{fig:relation} displays
typical size $a_{\infty}$ of condensed particles and supercooling
$\Delta T$ in terms of $\Lambda$ and $\Gamma$. 
The supercooling $\Delta T$ in the vertical axis is normalized by the equilibrium condensation temperature $T_{\rm e}$. Each solid curve shows the relation between $a_{\infty}$ and $\Delta T$ for a constant value of $\Gamma$, and dashed lines combine points for the same value in $\Lambda$ \citep{yamamoto77}. The grayed region indicates a parameter range expected from the planetesimal bow shock. The possible range of $\Lambda$ was discussed in \S \ref{sec:cooling_parameter}. The values of $\Gamma$, on the other hand, have uncertainties
because of short of the experimental data for the surface tension $\gamma_{\rm s}$ of forsterite and enstatite. For forsterite, $\gamma_{\rm
s}$ is measured to be $1280 \ {\rm erg \ cm^{-2}}$ in vacuum for a
\{010\} surface and larger values for other ones \citep{deleeuw00}, which corresponds to $\Gamma \simeq 30$ or more. For enstatite,
there are no reliable data of surface tension. We assume the similar
value as that of forsterite. In the calculations, we take $10 < \Gamma <
60$ for safety. It should be noted that the
sizes $a_{\infty}$ and the supercoolings $\Delta T$ revealed from the
analyses of a variety of cosmic crystals are included in the region
realized by planetesimal bow shocks.  Let us discuss in more detail the
formation conditions of each of the cosmic crystals shown in
Fig.~\ref{fig:relation}.

\subsubsection{Enstatite whisker and platelet}
The experiment by \citet{yamada09} showed that formation of enstatite
whiskers elongated toward the $a$-axis required the degree of
supercooling of $0.15 < \Delta T / T_{\rm e} < 0.48$. They also
reproduced platy-type enstatite crystals at $0.04 < \Delta T / T_{\rm e}
< 0.15$. It is interesting to note that the whisker has larger $\Gamma$
than the platy-type, although precise values of their surface tension
are unknown. Typical size of the enstatite crystals is $\sim 0.1$ - $1 \
{\rm \mu m}$, which size is similar to that of natural samples found in
IDPs \citep{bradley83}. A set of these conditions is shown by
the red region in Fig.~\ref{fig:relation}, indicating that enstatite
whiskers and platelets can be formed by planetesimal bow shocks of $10^3
\la \Lambda \la 10^4$. This range of $\Lambda$ is realized if the bow
shocks are produced by planetesimals of intermediate size ($R_{\rm p}
\sim 100 \ {\rm km}$) and lead to almost complete evaporation of the
original dust ($\eta \simeq 1$). If the amount of the silicate vapor is
small leaving a large amount of dust particles that survived evaporation
($\eta \ll 1$), on the other hand, the vapor will condense onto the dust
surface. This case yields other types of thermally-processed particles
observed in chondritic meteorites (see \S \ref{sec:discussion}).

\subsubsection{Olivine crystals with various morphologies}
\citet{kobatake08} examined supercooling $\Delta T$ required for
formation of olivine crystals by a laboratory experiment. They showed
that bulky-type olivine crystals were reproduced at $\Delta T / T_{\rm
e} \la 0.12$, the platy-type at $0.12 < \Delta T / T_{\rm e} < 0.33$,
and the columnar-needle-type at $0.24 < \Delta T / T_{\rm e} < 0.47$.
As was so for enstatite, the needle-type has larger $\Gamma$ than the
platy-type; the bulky type has the smallest $\Gamma$.  Typical size of
the condensates is $a_{\infty} \sim \mu$m, which size is close to those
of the natural samples found in the matrix of Allende meteorite \citep{nozawa09}.  The green region shows the supercooling $\Delta T /
T_{\rm e}$ and the sizes $a_{\infty}$ for these fine olivine crystals,
indicating that these particles can be formed by planetesimal bow shock
of $10^4 \la \Lambda \la 10^5$. This condition is realized by the bow
shocks produced by relatively large planetesimals ($R_{\rm p} \sim 1000
\ {\rm km}$) associated with almost complete evaporation of the original
dust particles.

\subsubsection{Ultra-fine particles}
\citet{toriumi89} observed fine particles in the matrix of Allende
meteorite using a SEM and a TEM and measured their sizes $a_{\rm
d}$. The observed size distribution could be reproduced by a log-normal
one for $1 < a_{\rm d} < 10 \ {\rm nm}$ with its peak at $a = 5 \ {\rm
nm}$ and by a power law for $a_{\rm d} > 10\ {\rm nm}$. We display the
size range of the ultra-fine particles by the blue region in
Fig.~\ref{fig:relation}.  The size range suggests that $\Lambda \simeq
10-100$ is a plausible condition for formation of ultra-fine
particles. This is in agreement with the conclusion of \citet{toriumi89}
that ultra-fine particles seem to have been formed by condensation from
a vapor far from equilibrium in the early solar nebula. The present
model implies that ultra-fine particles were formed by bow shocks produced by much smaller planetesimals ($R_{\rm p} \sim 1 - 10 \ {\rm km}$) than those producing $\mu$m-sized cosmic crystals, associated with almost complete evaporation. The formation condition of $\Lambda \simeq 10-100$ also realizes for large planetesimals ($R_{\rm p} \ga 100 \ {\rm km}$) and small evaporation fraction ($\eta \sim 10^{-4} - 10^{-3}$), however, in this case the ultra-fine particles generated from the vapor are very rare because of the tiny evaporation fraction.

\subsection{Heterogeneous condensation for incomplete evaporation}
\label{sec:homogeneous_vs_heterogeneous}
We discussed formation of cosmic crystals through homogeneous nucleation in \S \ref{sec:cooling_parameter} and \S \ref{sec:averagedsize_condensationtemperature} assuming that almost all dust particles evaporate by a planetesimal bow shock.
There is an opposite case that a substantial fraction of the dust particles survives against evaporation and acts as seed nuclei and that condensation occurs through nucleation on their surfaces (heterogeneous condensation).
Which type of
condensation actually occurs depends on the total surface area of dust
particles available for heterogeneous nucleation. We shall show below
that both types of condensations can occur depending on the radii of
planetesimals generating bow shocks and on the evaporation fraction.

In homogeneous nucleation, condensation does not begin when the cooling
vapor become saturated but begins effectively after the vapor becomes
supersaturated to a certain degree. Namely, there arises some induction
time $t_{\rm ind}$ after the vapor becomes saturated \citep{yamamoto77}. The induction time is related to the cooling timescale
$t_{T} = T_{\rm e} / (-dT/dt)_{T_{\rm e}}$ as
\begin{equation}
t_{\rm ind} \simeq \frac{ x_{\rm J} }{ H/T_{\rm e} - 1 } t_{T} 
\sim (0.08 - 4.0)
 \left( \frac{ R_{\rm p} }{ 1 \ {\rm km} } \right) \ {\rm s} ,
\label{eq:homo}
\end{equation}
where $x_{\rm J} = 2 - 70$ for situations we consider in this paper ($\Lambda = 1 - 10^5$ and
$\Gamma = 10 - 60$ as explained in \S\ref{sec:cooling_parameter}
and \S \ref{sec:averagedsize_condensationtemperature}). The time intervals required for nucleation and growth is
about ten times shorter than the induction time \citep{yamamoto77}. Therefore, $t_{\rm ind}$ represents a typical timescale for dust
formation through homogeneous nucleation after the vapor becomes
saturated. 

In heterogeneous nucleation, on the other hand, we estimate its
timescale by using the adhesion timescale, which provides an
underestimate of the timescale of heterogeneous condensation because it
ignores the induction time for heterogeneous nucleation. The adhesion
timescale $t_{\rm ad}$ is the one during which most of the vapor
molecules sticks onto the surface of dust particles. For silicate
condensation, we regard SiO molecule as a key species that controls the
rate of condensation (see also Appendix \ref{sec:multi_component}). Denoting
the radius of the dust particles by $a_{\rm d}$, the adhesion timescale
is estimated to be:
\begin{equation}
t_{\rm ad}
= \frac{ a_{\rm d} \rho_{\rm c} }{ 3 \alpha_{\rm s} \rho_{\rm d} }
\left( \frac{ 2 \pi \mu_{\rm SiO} m_{\rm a} }
            { k_{\rm B} T_{\rm e} } \right)^{1/2}
\simeq
50 \ \alpha_{\rm s}^{-1} (1 - \eta)^{-1}
\left( \frac{ a_{\rm d} }{ {\rm \mu m} } \right)
\left( \frac{ 10^{-10} \ {\rm g \ cm^{-3}} }
            { \rho_{\rm d} + \rho_{\rm v} } \right)\ {\rm s} ,
\label{eq:hetero2}
\end{equation}
where $\rho_{\rm d} = (1 - \eta) (\rho_{\rm d} + \rho_{\rm v})$ is
density of dust particles surviving in the post-shock region against
evaporation, $\alpha_{\rm s}$ is sticking probability of vapor molecules
onto the dust surface, and $\mu_{\rm SiO} = 44$ is molecular weight of SiO. One
should note that, in Eq. (\ref{eq:hetero2}), the factor $3 \rho_{\rm d}
/ \rho_{\rm c} a_{\rm d}$ indicates total surface area of the dust
particles per unit volume and $(8 k_{\rm B} T_{\rm e} / \pi \mu_{\rm
SiO} m_{\rm a})^{1/2}$ is mean thermal velocity of SiO molecules.

Homogeneous nucleation takes place if $t_{\rm ind} < t_{\rm ad}$. This
condition is satisfied when the planetesimal radius $R_{\rm p}$ is
relatively small ($R_{\rm p} \la 10-500 \ {\rm km}$), or there are few
survived dust particles because of significant evaporation ($\eta \sim
1$). In this case, cosmic crystals condense directly from the vapor.  In
contrast, heterogeneous condensation becomes effective if the
planetesimal radius $R_{\rm p}$ is large ($R_{\rm p} \ga 10-500 \ {\rm km}$)
{\em and} a substantial fraction of dust particles survives against
evaporation ($\eta \ll 1$).  We shall discuss generic relations between
cosmic crystals and chondrules in \S~\ref{sec:discussion} in detail.

\section{Summary and Discussion}
\label{sec:discussion}
Chondritic meteorites are composed of materials that have been
experienced thermal processing of various degrees in the early solar
nebula.  These materials include chondrules, fine-grained rims on
chondrules and interchondrule matrix \citep{alexander95}, and cosmic
crystals discussed in the previous section.
In this section, we discuss how the planetesimal bow shock scenario
explains the formations of these chondritic materials.

A planetesimal bow shock was originally proposed as a possible site for
chondrule formation \citep{hood98}. \cite{iida01} showed that millimeter-sized dust aggregates (chondrule precursors) are heated and melt behind a shock front if the shock velocity and the pre-shock gas density are in an appropriate range. Complete evaporation
hardly occurs for chondrule precursors because of their large size (see
Eq.~(\ref{eq:evaporation_fraction-text})). Their contribution to the
vapor production is negligibly small compared with that of $\mu$m-sized
dust particles for the dust size distribution steeper than $a_{\rm
d}^{-2}$. Large molten chondrule precursor dust survives against
evaporation, cools and solidifies to form chondrules.  In contrast,
$\mu$m-sized particles evaporate significantly in the hot gas behind the
planetesimal bow shock and produce silicate vapor.
The silicate vapor cools rapidly behind the bow shock and becomes
supersaturated, leading to condensation to produce various kinds of
materials observed in chondritic meteorites and IDPs.

The condensed materials exhibit a wide variety in morphologies and sizes
depending on their formation conditions such as the cooling rate and the
evaporation fraction of $\mu$m-sized dust particles. The cooling rate is
inversely proportional to the size of a planetesimal that produces a bow
shock (see Eq.~(\ref{-dT/dt2})), thus decreases with time on average,
namely, with growth of planetesimals.  The evaporation fraction $\eta$
changes by many orders of magnitude in the range of the shock conditions
realized in early solar nebula \citep{iida01}.

Figure~\ref{fig:summary} summarizes condensation products in the
course of the planetesimal growth.  At the early stage of $1 \la R_{\rm
p} \la 10\ $km, the vapor produced by small planetesimals cools so
rapidly that the cooling parameter is $\Lambda \la 10^3$, which is
realized for $10^{-2} \la \eta \la 1$ (see Fig.~\ref{fig:Lambda}b).
Condensation of the vapor through homogeneous nucleation for $1 <
\Lambda \la 10^3$ leads to formation of nm-sized ultra-fine particles as
observed in the matrix (see from Fig.~\ref{fig:relation}).  Furthermore,
the results of \S\ref{sec:homogeneous_vs_heterogeneous} indicate that
the rapid cooling prevents heterogeneous condensation on survived dust
particles because of rapid consumption of the vapor by homogeneous
condensation to form ultra-fine particles. To summarize, most of the
vapor condensed to the ultra-fine particles at the early stage of
planetesimal growth. When planetesimals grow up to a several $100 \ {\rm km}$ or
more, condensation occurs through both homogeneous and heterogeneous
nucleations. If almost all of the small dust particles evaporate ($\eta
\sim 1$) behind the bow shock, $\mu$m-sized euhedral silicate crystals
condense through homogeneous nucleation. The cooling parameter is $10^3
\la \Lambda \la 10^5$ for $\eta \sim 1$ (see Fig.~\ref{fig:Lambda}b) for
planetesimals of $100 \la R_{\rm p} \la 1000\ $km. This situation leads
to condensation of enstatite whisker elongated to $a$-axis as found in
IDPs \citep{bradley83} and polyhedral olivine crystals as found in the
matrix of Allende meteorite \citep{nozawa09} (see
Fig.~\ref{fig:relation}). Bare chondrules without fine-grained rims
could also be formed in this case. 
On the other hand, if many of the dust particles survive against
evaporation ($\eta \ll 1$) and suffer partial evaporation, the vapor
condenses heterogeneously onto the survived dust particles, resulting in
the formation of other kinds of meteoritic materials. The vapor
condensed heterogeneously on chondrules already solidified could form
fine-grained rim on their surfaces. The survived $\mu$m-sized dust
particles would also be covered with materials condensed from vapor, and
would accumulate as fine-grained interchondrule matrix in chondritic
meteorites after that. It is worth noting that the partial evaporation
of dust particles would lead to elemental fractionation.  The
fractionated vapor rich in volatile elements re-condensed within a short
period of time (see Eq.~(\ref{eq:hetero2})) on the survived dust
particles.  A fine-grained rim of a chondrule and a fine-grained
interchondrule matrix thus produced would have elemental composition
complementary to that of the chondrule. The composition of the whole
particle should be the same as that of the original dust particles
before evaporation according to the present model. This is consistent
with the genetic relationship among chondrules, interchondrule matrix,
and fine-grained rims that these components either formed from a common
source material, are products of the same process, or have exchanged
materials during formation \citep{huss05}.

In summary, the planetesimal bow shock model can provide a comprehensive
scenario for the formation of various cosmic crystals and other
materials observed in chondritic meteorites. Because the heating events
happened in a localized region of the shocked gas within a short period
of time, one may expect that a series of the thermal processing,
heating, evaporation, and condensation completed in a closed-system. The
scenario is in harmony with the genetic relationship suggested by the
analyses of chondritic meteorites and IDPs that these are produced in
the course of the processing from a common source material together with
exchanges of the materials during their formation.

\acknowledgments

We are grateful to Dr. M. Nagasawa for useful discussion on the orbital evolutions of planetesimals. We acknowledge helpful comments of an anonymous referee. This study was supported partly by the Grant for the Joint Research Program of the Institute of Low Temperature Science, Hokkaido University. H.M. was supported by Tohoku University Global COE Program ``Global Education and Research Center for Earth and Planetary Dynamics," by the ``Program Research'' in Center for Interdisciplinary Research, Tohoku University, and by the Grant-in-Aid for Scientific Research from JSPS (19204052). T.Y. acknowledges support by the Grant-in-Aid for Scientific Research from JSPS (21244011).


%% in parentheses, after the keyword, but they will not be verified.

%{\it Facilities:} \facility{Nickel}, \facility{HST (STIS)}, \facility{CXO (ASIS)}.



\appendix

\section{Size Decrease of a Particle by Evaporation
 and the Evaporation Fraction}
\label{sec:analytic_formula_evaporation_fraction}
Let us consider evaporation of a spherical dust particle of
initial radius $a_{\rm d}$. The evaporation fraction $\eta$ is given
by 
\begin{equation}
\eta
= \frac{3}{4 \pi a_{\rm d}^3}
\int_{a_{\rm d} - \Delta a}^{a_{\rm d}} 4 \pi a^2 da
= 1 - \left( 1 - \frac{ \Delta a }{ a_{\rm d} } \right)^3 ,
\label{eq:evaporation_fraction}
\end{equation}
where $\Delta a$ is decrease in radius due to evaporation. Equation
(\ref{eq:evaporation_fraction}) indicates clearly that $\eta$ depends
only on $\Delta a / a_{\rm d}$, the ratio of the size decrease to the
initial size. Since the size decrease $\Delta a$ due to evaporation is
independent of the particle radius $a_{\rm d}$ except through a slight
dependence of the dust temperature on $a_{\rm d}$, the increase in the
particle radius $a_{\rm d}$ simply causes the decrease in $\eta$
according to Eq.~(\ref{eq:evaporation_fraction}). For example, even if
$\mu$m-sized dust particles evaporate almost completely ($\eta =
0.999$), chondrule-sized particles ($a_{\rm d} = 500 \ {\rm \mu m}$)
evaporate by only a small fraction of $\eta \simeq 5\times10^{-3}$.  We
carry out the calculations for $a_{\rm d} = 1 \ {\rm \mu m}$, but one
can evaluate $\eta$ for other $a_{\rm d}$ by using
Eq.~(\ref{eq:evaporation_fraction}).

With the use of the evaporation rate $j_{\rm evap}$, the size decrease
$\Delta a$ by evaporation during cooling from temperature $T_{\rm i}$ to $T_f$
is expressed by
\begin{equation}
\Delta a
= \frac{1}{ \rho_{\rm c} } \int j_{\rm evap} dt
= \frac{1}{ \rho_{\rm c} }
\int_{T_{\rm i}}^{T_{\rm f}} \frac{ j_{\rm evap} (T) }{dT/dt} dT.
\label{deltaaint}
\end{equation}
The evaporation rate as a function of temperature $T$ behaves as
\begin{equation}
j_{\rm evap} (T) = {\rm const} \cdot T^{\beta} \exp(-H/T) ,
\end{equation}
according to the Hertz-Knudsen equation \citep[see]{nagahara96, miura02}, where $\beta \sim -1/2$ is a constant. 
Here, $H = L_{\rm evap} / R_{\rm gas}$ is latent heat of evaporation in units of temperature, where $L_{\rm evap}$ is that in units ${\rm J \ mol^{-1}}$ and $R_{\rm gas}$ is the gas constant in units of ${\rm J \ K^{-1} \ mol^{-1}}$. In the present case, we are concerned with evaporation of forsterite (Mg$_2$SiO$_4$), for which $L_{\rm evap} = 1.58\times10^{13} \ {\rm J \ mol^{-1}}$. This leads $H = L_{\rm evap} / 6 R_{\rm gas} = 3.17\times10^4 \ {\rm K}$, where the factor of 6 results from the stoichiometric coefficients of the chemical reactions at evaporation \citep[see Eq. (38)]{miura02}.
According to the measurement of evaporation rate of forsterite, $j_{\rm evap}$ depends also
on the partial pressure $p_{\rm H_2}$ of ambient hydrogen molecule \citep{tsuchiyama98}, but
we may take the pressure at $T = T_{\rm i}$ in Eq.~(\ref{deltaaint})
because $\Delta a$ is determined by the physical conditions at $T =
T_{\rm i}$ as will be seen below.
Integration on the RHS of Eq.~(\ref{deltaaint}) can be performed by
noting that $e^{-H/T}$ is a rapidly varying function compared to the
remaining function in the integrand.  Integrating by part and remaining
the term of order $T_{\rm i}/H \ll 1$, one obtains
\begin{equation}
\Delta a
= \frac{ j_{\rm evap} (T_{\rm i}) }{ \rho_{\rm c} }
\frac{ T_{\rm i} }{ H } \Delta t , 
\label{deltaa-fin}
\end{equation}
where the contribution from the upper limit of the integral is
negligible.  Here
\begin{equation}
\Delta t = \frac{T_{\rm i}}{(-dT/dt)_{T_{\rm i}} }
\label{Delta-t}
\end{equation}
is cooling timescale of dust particles at $T = T_{\rm i}$, for which we take
dust temperature $T_{\rm d2}$ in the second stage.

\section{Dust Temperature behind Shock Front}
\label{sec:thermal_conduction}
We give here an analytic expression that gives in good approximation of
the dust temperature in the post-shock region after the relative
velocity between the gas and dust particles is almost damped (the second
stage). The dust temperature at this stage, $T_{\rm d2}$, is determined
by the energy balance between collisional heating by the ambient hot gas
and the radiative cooling:
\begin{equation}
\frac{1}{4} \frac{ \gamma + 1 }{ \gamma - 1 } 
\left( \frac{ \gamma k_{\rm B} T' }{ \pi \mu m_{\rm a} } \right)^{1/2} n' k_{\rm B}
\left( T' - T_{\rm d2} \right)
+ \sigma_{\rm SB} (T_{\rm rad}^4 - T_{\rm d2}^4)
= 0 ,
\label{eq:balance_smallsa}
\end{equation}
where $T'$ is post-shock gas temperature, $n'$ is post-shock gas number
density, $T_{\rm rad}$ is ambient radiation temperature, $\mu$ is mean
molecular weight of the gas, and $\sigma_{\rm SB}$ is the
Stafan-Boltzmann constant.  Here, we approximated the emission and
absorption coefficients to be unity \citep{miyake93}.
Since $T_{\rm rad} \sim T_{\rm d2}$ in the second stage as seen from
Fig.~\ref{fig:profile_01}(c), one obtains
\begin{equation}
T_{\rm d2}
= \frac{ \gamma_r ( \gamma k_{\rm B} T' / \pi \mu m_{\rm a} )^{1/2}
 n' k_{\rm B} T'/4 + 4 \sigma_{\rm SB} T_{\rm rad}^4 }
{ \gamma_r ( \gamma k_{\rm B} T' / \pi \mu m_{\rm a} )^{1/2}
 n' k_{\rm B} / 4 + 4 \sigma_{\rm SB} T_{\rm rad}^3 } ,
\label{eq:dust_temp_cond}
\end{equation}
from Eq.~(\ref{eq:balance_smallsa}) by using the approximation that
$(T_{\rm rad}^4 - T_{\rm d2}^4) \simeq 4 T_{\rm rad}^3 (T_{\rm rad}
- T_{\rm d2})$, where $\gamma_r \equiv (\gamma+1)/(\gamma-1)$. 

In Eq.~(\ref{eq:dust_temp_cond}), the post-shock gas number density $n'$
is given from the Rankine-Hugoniot relation and the almost isobaric
condition for the post-shock gas in the one-dimensional plane-parallel
geometry \citep{susa98, miura02} by
\begin{equation}
n' \simeq \frac{ 2 }{ \gamma + 1 }
        \frac{\rho_0 v_{\rm s}^2}{ k_{\rm B} T' },
\label{n'}
\end{equation}
where $\rho_0 = \mu m_{\rm a} n_0$ is the gas density in the pre-shock
region and $v_{\rm s}$ is the shock velocity.

The post-shock gas temperature $T'$ in Eq.~(\ref{eq:dust_temp_cond}) may
be obtained by using the Rankine-Hugoniot relation. However, we have to
pay attention that, at high temperatures of $T' \ga 2000 \ {\rm K}$, the
gas cools very rapidly due to the dissociation of hydrogen molecules
\citep{iida01}. Therefore, we set $T'$ as
\begin{equation}
T' = {\rm min} \left[ \frac{ 2 (\gamma-1) }
   { ( \gamma + 1 )^2 } \frac{ \mu m_{\rm a} v_{\rm s}^2}{ k_{\rm B} },
    ~ 2000 \ {\rm K} \right] .
\label{T'}
\end{equation}
The radiation temperature at the shock front is given by taking the
blanket effect into account \citep{miura06} as
\begin{equation}
T_{\rm rad}
= \left( \frac{ 2 + 3 \tau_{\rm pre} }{ 4 \sigma_{\rm SB} } 
 \frac{f}{2} \rho_0 v_{\rm s}^3 \right)^{1/4} ,
\label{eq:radiation_temperature}
\end{equation}
where $\tau_{\rm pre}$ is optical depth of the pre-shock region and $f$
is the fraction of the gas energy flux that returns upstream in the form
of radiation. We set $f = 0.5$ for simplicity. The optical depth
$\tau_{\rm pre}$ is estimated to be 
\begin{equation}
\tau_{\rm pre}
= \frac{ 3 \xi \rho_0 }
      { 4 a_{\rm d} \rho_{\rm c} } L_{\rm s} ,
\end{equation}
where $\xi$ is dust-to-gas mass ratio in the pre-shock region and
$L_{\rm s}$ is dimension of the pre-shock region, in which the dust
particles contribute to the blanket effect around the shock
front\footnote{$L_{\rm s}$ corresponds to $x_{\rm m}$ in Miura \&
Nakamoto (2006).}. We set $L_{\rm s} = R_{\rm p}$ in this study
(see \S \ref{sec:numerical_simulation}).

Equation~(\ref{eq:dust_temp_cond}) together with Eqs.~(\ref{n'}),
(\ref{T'}), and (\ref{eq:radiation_temperature}) is an analytic
expression of the dust temperature in the second stage. It should be
noted that these equations include all of the input parameters $L_{\rm
s} = R_{\rm p}$, $n_0 = \rho_0/\mu m_{\rm a}$, $\xi$, and $v_{\rm s}$.

\section{Solutions of the Equations of Expansion}
\label{sec:analytic_solution}

\begin{figure}
\epsscale{.70}
\plotone{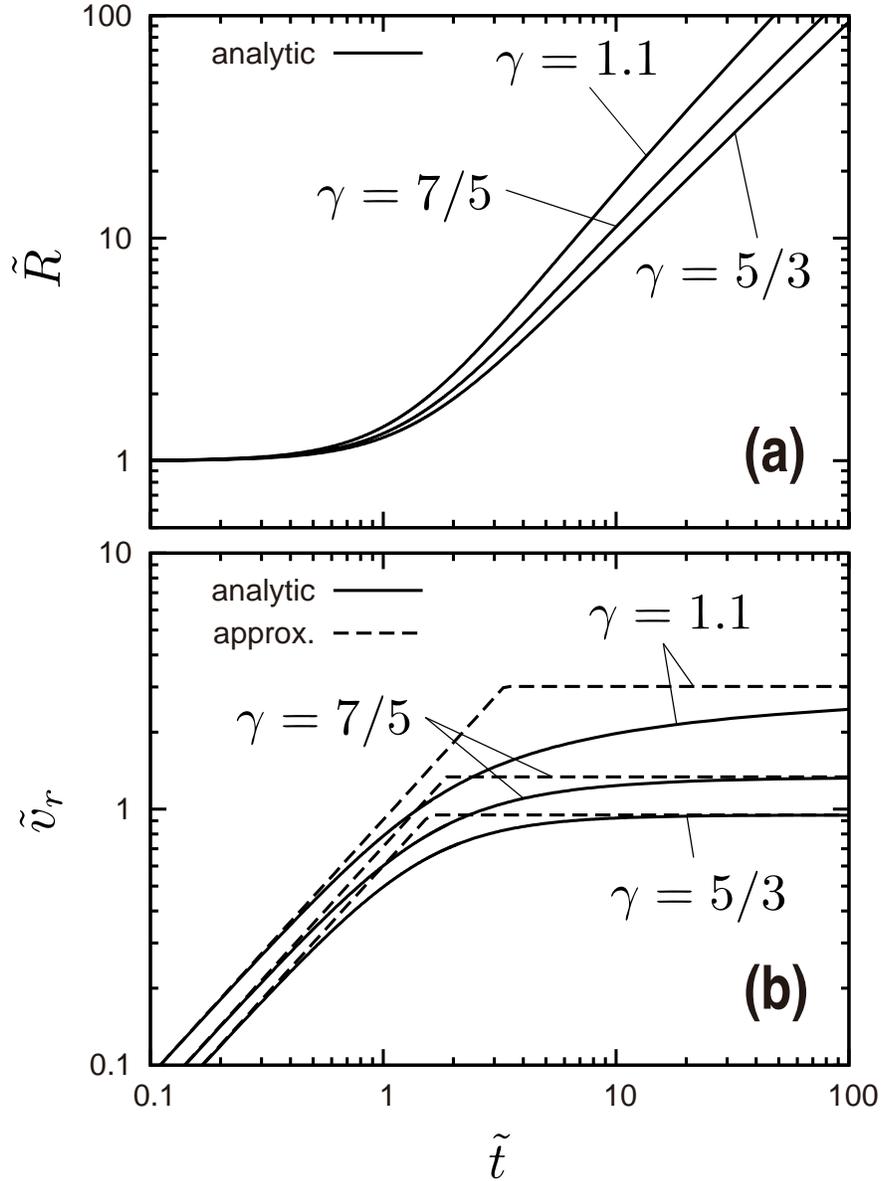}
\caption{%
Temporal variations of (a)~the radius $\tilde{R}$ of the gas cloud and
(b)~its expansion velocity $\tilde{v}_r$ behind a planetesimal bow
shock. The curves for $\gamma = 7/5$ correspond to adiabatic expansion
of a gas composed of ${\rm H_2}$ molecules. All quantities including
time $\tilde{t}$ are normalized (see text for details). The solid curves
show exact solutions given by Eq.~(\ref{eq:radius}) for $R$ and by
Eq.~(\ref{eq:vr_solution}) for $v_r$, while the dashed curves show
approximations in the two limiting cases given by Eq.~(\ref{vr-limits}).
}
\label{fig:expansion}
\end{figure}

\begin{figure}
\epsscale{.80}
\plotone{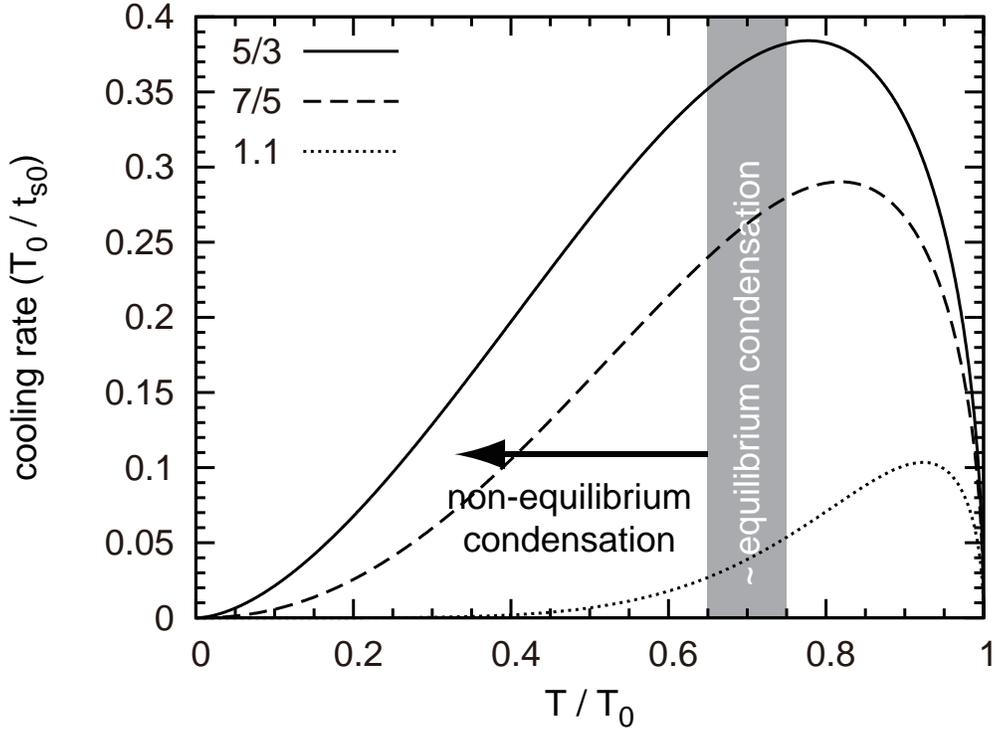}
\caption{%
 Cooling rate of the vapor produced by a planetesimal bow shock during
 its expansion versus the gas temperature $T$. The horizontal axis is
 $T$ normalized by the initial temperature $T_0$ and the vertical one is
 the cooling rate normalized by $T_0/t_{\rm s0}$, where $t_{\rm s0} =
 R_0/c_{\rm s0}$ is the sound-crossing time. The solid, dashed, and
 dotted curves show cooling rates for $\gamma = 5/3$, $7/5$, and $1.1$,
 respectively. The gray region indicates a range of the equilibrium
 condensation temperatures of silicates under the total pressure of protoplanetary disk, $T_{\rm e} = 1300 - 1500 \ {\rm
 K}$.
}
\label{fig:Rcool_temp}
\end{figure}

\begin{figure}
\epsscale{.80} \plotone{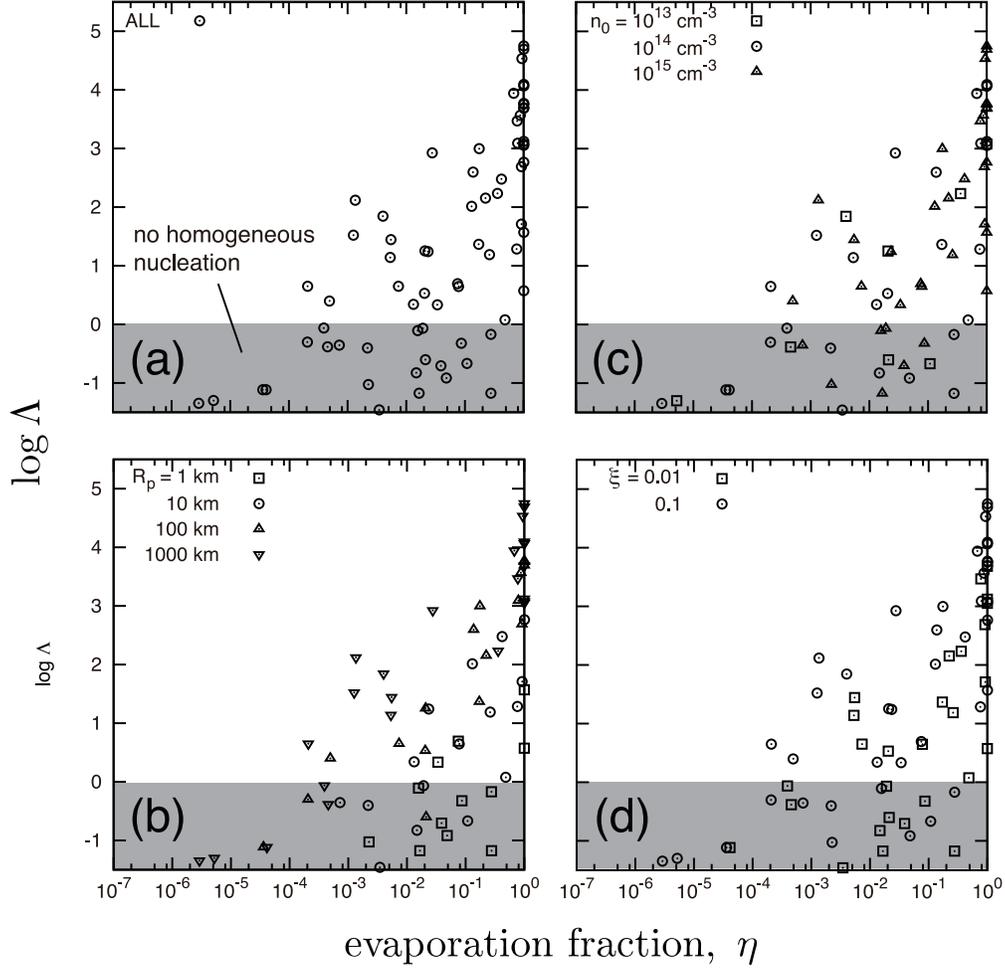} 
\caption{ Relation between the cooling parameter $\Lambda$ and the evaporation fraction
$\eta$. Plotted are $\eta$ calculated by the numerical simulations and
$\Lambda$ calculated from Eq.~(\ref{eq:lambda}): (a) for the data for
all all of the parameter sets, (b) for each $R_{\rm p}$, (c) for each
$n_0$, and (d) for each $\xi$. The gray region ($\Lambda < 1$)
indicates the region where condensation through homogeneous nucleation
does not take place during the vapor cooling.
}
\label{fig:Lambda}
\end{figure}

\begin{figure}
\epsscale{.80}
\plotone{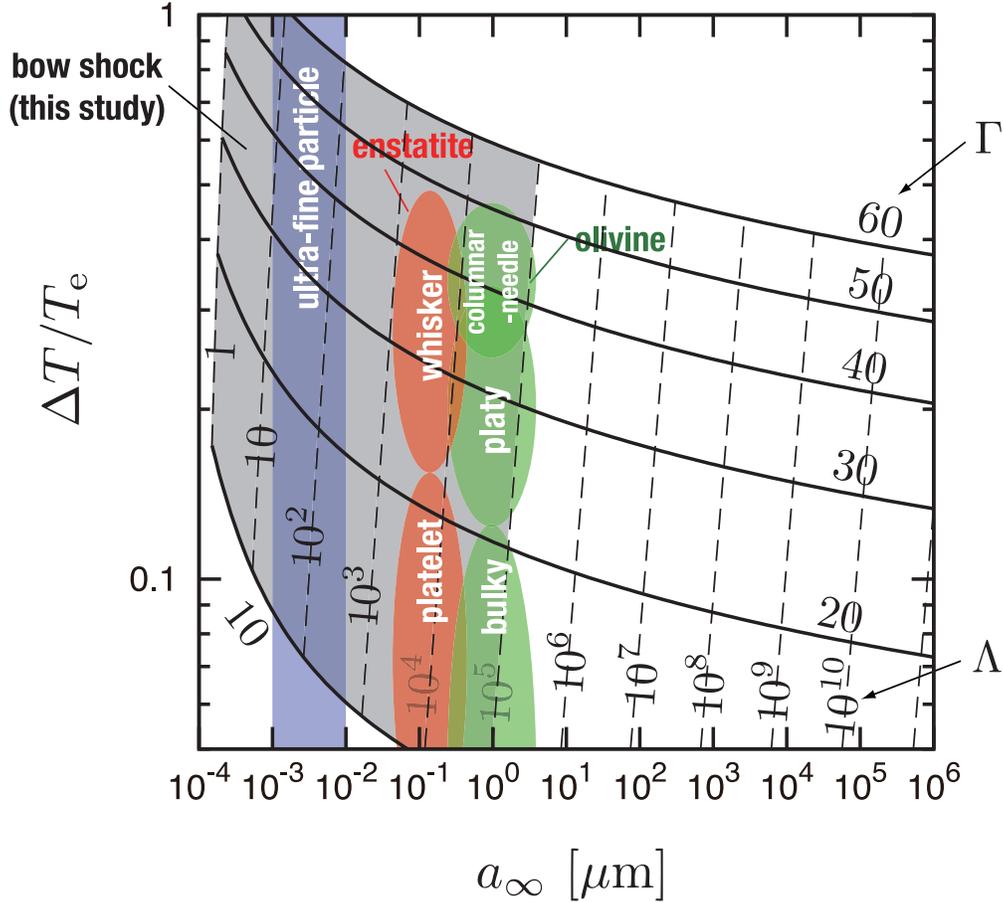}
\caption{ 
Typical radii $a_{\infty}$ of the particles condensed in the vapor and
the degree of the supercooling $\Delta T/T_{\rm e}$ versus the cooling
parameter $\Lambda$ and the dimensionless surface tension $\Gamma$. The
gray region indicates the ranges of $\Lambda$ expected from the model
and of possible values of $\Gamma$ of the condensates. Solid
curves indicate $a_{\infty}$ and $\Delta T$ calculated based on the
homogeneous nucleation theory for $\Gamma = 10$ to 60. Dashed lines
indicate those for $\Lambda = 1$ to $10^{10}$. Products of the
evaporation and condensation experiments are shown by the red (enstatite
crystals, Yamada 2009) and green regions (forsterite crystals, Kobatake
et al., 2008). The typical size range of ultra-fine particles in the
matrix of Allende meteorite (Toriumi, 1989) is shown by the blue region.
} 
\label{fig:relation}
\end{figure}

\begin{figure}
\epsscale{.80}
\plotone{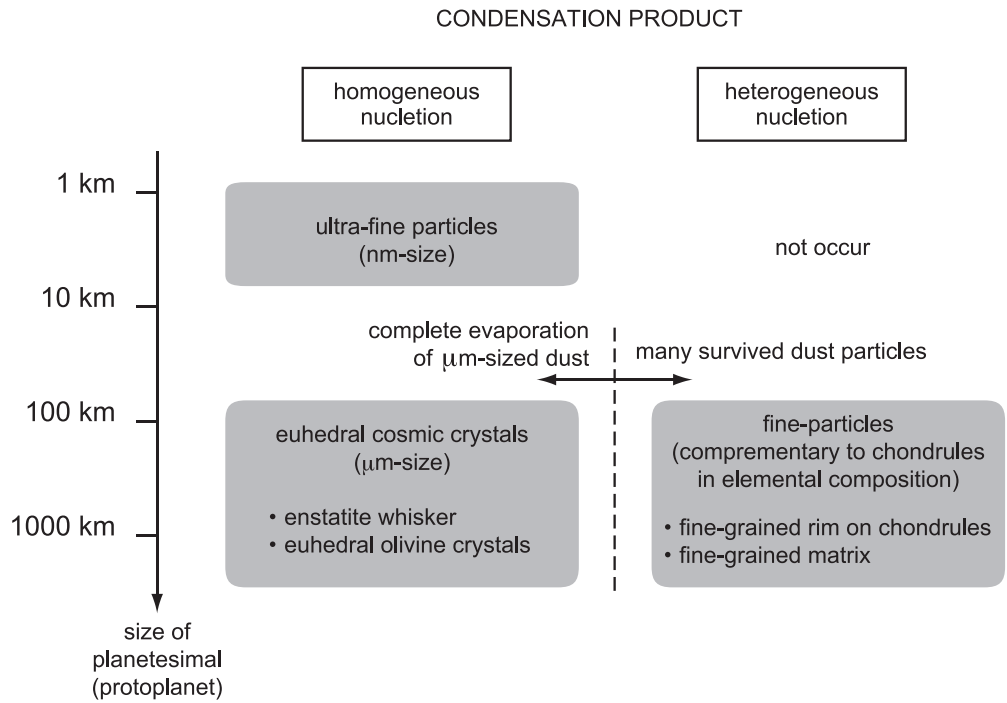}
\caption{
 Formation of chondritic materials produced by planetesimal bow shocks
 in the course of planetesimal growth.
}
\label{fig:summary}
\end{figure}

Integrating Eq.~(\ref{eq:eom_r02}) from $\tilde{t} = 0$ to $\tilde{t}$
after multiplying $d\tilde{R}/d\tilde{t} = \tilde{v}_{r}$ on both sides,
we obtain the expansion velocity to be:
\begin{equation}
\tilde{v}_r
= \frac{d\tilde{R}}{d\tilde{t}}
= \left[ \frac{1 - \tilde{R}^{-2(\gamma-1)}}
            {\gamma(\gamma-1)} \right]^{1/2} .
\label{eq:vr_solution}
\end{equation}
for the initial conditions of $\tilde{v}_{r} = 0$ and $\tilde{R} = 1$.
Equation~(\ref{eq:vr_solution}) is integrated further to yield the
radius $\tilde{R}$ as a function of time $\tilde{t}$ as
\begin{equation}
\frac{ \tilde{t} }{ \sqrt{\gamma(\gamma-1)} } 
= 
\int_1^{\tilde{R}} \frac{dy}{\sqrt{1 - y^{-2(\gamma-1)}}}.
\label{eq:radius}
\end{equation}
The right-hand side of Eq. (\ref{eq:radius}) may be expressed by the
hypergeometric function but numerical integration is more practical to
get the results, which are shown in Fig.~\ref{fig:expansion}(b) by solid
curves.

In the limits of $\tilde{t} \ll 1$ and $\tilde{t} \rightarrow \infty$,
the velocity is approximated to be:
\begin{equation} 
\tilde{v}_r = \left\{
\begin{array}{ll}
\displaystyle{ \frac{ \tilde{t} }{\gamma} } & (\tilde{t} \ll 1), \\
\displaystyle{ \left[\frac{1}{\gamma (\gamma - 1)}\right]^{1/2} } & (\tilde{t} \rightarrow \infty). 
\label{eq:expansion_app}
\end{array}
\right.
\end{equation}
Both approximations are shown in Fig. \ref{fig:expansion}(b) by the
dashed lines. The time at the intersection $\tilde{t}_{*}$, at which the
two limiting approximations cross each other, is given by
\begin{equation}
\tilde{t}_{*} = \left( \frac{ \gamma }{ \gamma - 1 } \right)^{1/2} .
\end{equation}
At the intersection, the radius and expansion velocity are given by
\begin{equation}
\tilde{R}_{*} = \frac{ 2 \gamma - 1 }{ 2 (\gamma-1) } , 
\quad
\tilde{v}_{r*} = \left[ \frac{ 1 }{ \gamma (\gamma-1) } \right]^{1/2} .
\label{eq:sol_intersection}
\end{equation}

\section{Evaluation of cooling parameter $\Lambda$ for multi-component evaporation}
\label{sec:multi_component}
In this paper, we are concerned with evaporation and condensation of forsterite (Mg$_2$SiO$_4$), in which Mg and SiO should be considered as vapor species \citep{nagahara96}. For dealing with nucleation of a multi-component system, we adopt the key species approximation that the rates of nucleation and grain growth are controlled by one chemical species (key species) that has the least collision frequency among the major vapor species that condense into the grain \citep{kozasa87}. The conditions for the key-species approximation to hold were examined by \citet{yamamoto01} in formulating theory of nucleation involving chemical reactions.

In Eq.~(\ref{eq:lambda01}), the collision
frequency of vapor molecules of mass $\mu_{\rm v} m_{\rm a}$ and number
density $n_{\rm v}$ is given by $\nu_{\rm coll} = \pi a_0^2 \alpha_{\rm
s} n_{\rm v} \sqrt{ 8 k_{\rm B} T_{\rm e}/ \pi \mu_{\rm v} m_{\rm a}}$,
where $\alpha_{\rm s}$ is sticking probability, $\mu_{\rm v}$ is mean
molecular weight of the vapor molecules. 
Following \citet{kozasa87}, we take SiO molecules as the key species of silicate condensation. This
implies that $n_{\rm v} = n_{\rm SiO}$ and $\mu_{\rm v} = \mu_{\rm SiO}
= 44$.  Using Eqs.~(\ref{-dT/dt2}) and (\ref{cs0}), we obtain
\begin{equation}
\Lambda
= 
 \frac{4 \pi a_0^2 \alpha_{\rm s} 
  n_{\rm SiO} R_{\rm p} }{ H / T_{\rm e} - 1 }
 \left( \frac{ \mu }{ \mu_{\rm SiO} } \right)^{1/2}
 \left( \frac{ T_{\rm e} }{ T_0 } \right)^{3/2}
 \left( \frac{ 1 }{ \pi \gamma } \right)^{1/2}
 \times \frac{ 1 }{ (0.25 - 0.35) } ,
\end{equation}
where $\mu$ is mean molecular weight of the gas.  The value of $H/T_{\rm
e}$ is estimated to be $H/T_{\rm e} - 1 \simeq 20$ for $H \simeq 3
\times 10^4\,$K for forsterite; enstatite yields the similar value. The
value of $a_0$ is given by $a_0 = 2.6 \ {\rm \AA}$. In consequence,
$\Lambda$ is evaluated roughly to be:
\begin{equation}
\Lambda \simeq 400 
 \left( \frac{ R_{\rm p} }{ 100 \ {\rm km} } \right)
 \left( \frac{ \rho_{\rm v} }{ 10^{-10} \ {\rm g \ cm^{-3}} } \right) 
\label{eq:lambda}
\end{equation}
for $\alpha_{\rm s} = 1$.




\clearpage

%% the file, uncomment the \plotone, \plottwo, and
%% If you need a layout that cannot be achieved with \plotone or
%% \plottwo, you can invoke the graphicx package directly with the
%% \includegraphics command or use \plotfiddle. For more information,
%% please see the tutorial on "Using Electronic Art with AASTeX" in the
%% documentation section at the AASTeX Web site,
%% http://www.journals.uchicago.edu/AAS/AASTeX.
%%
%% The examples below also include sample markup for submission of
%% supplemental electronic materials. As always, be sure to check
%% the instructions to authors for the journal you are submitting to
%% for specific submissions guidelines as they vary from

%% This example uses \plotone to include an EPS file scaled to
%% 80% of its natural size with \epsscale. Its caption
%% has been written to indicate that additional figure parts will be
%% available in the electronic journal.

%%%%%

%%%%%

%%%%%

%%%%%

%%%%%

%%%%%

%%%%%

%%%%%

\clearpage

\clearpage


\begin{thebibliography}{}
\bibitem[Alexander(1995)]{alexander95} Alexander, C. M. O'D. 1995, \gca, 59, 3247
\bibitem[Bradley et al.(1983)]{bradley83} Bradley, J. P., Brownlee, D. E., \& Veblen, D. R. 1983, Nature, 301, 473
\bibitem[Ciesla et al.(2004)]{ciesla04} Ciesla, F. J., Hood, L. L., Weidenschilling, S. J. 2004, Meteorit. Planet. Sci., 39, 1809
\bibitem[de Leeuw et al.(2000)]{deleeuw00} de Leeuw, N. H., Parker, S. C., Catlow, C. R. A., \& Price G. D. 2000, Phys. Chem. Minerals, 27, 332
\bibitem[Draine \& Salpeter(1977)]{draine77} Draine, B. T., \& Salpeter, E. E. 1977, J. Chem. Phys., 67, 2230
\bibitem[Grossman(1972)]{grossman72} Grossman, L. 1972, \gca, 36, 597
\bibitem[Hayashi et al.(1985)]{hayashi85} Hayashi, C. K., Nakazawa, K., \& Nakagawa, Y. 1985, in Formation of the solar system, ed. D. C. Black \& M. S. Matthews, M.S. (Univ. of Arizona Press, Tucson), 1100
\bibitem[Hewins et al.(2005)]{hewins05} Hewins, R. H., Connolly, Jr., H. C., Lofgren, G. E., \& Libourel, G. 2005, in Chondrites and the Protoplanetary Disk, ed. A. N. Krot, E. R. D. Scott, \& B. Reipurth (San Francisco: Astronomical Society of the Pacific), 286
\bibitem[Hood(1998)]{hood98} Hood, L. L. 1998, Meteorit. Planet. Sci., 33, 97
\bibitem[Hood et al.(2005)]{hood05} Hood, L. L., Ciesla, F. J., \& Weidenschilling, S. J. 2005, in Chondrites and the Protoplanetary Disk, ed. A. N. Krot, E. R. D. Scott, \& B. Reipurth (San Francisco: Astronomical Society of the Pacific), 873
\bibitem[Hood et al.(2009)]{hood09} Hood, L. L., Ciesla, F. J., Artemieva, N. A., Marzari, F., \& Weidenschilling, S. J. 2009, Meteorit. Planet. Sci., 44, 327
\bibitem[Huss et al.(2005)]{huss05} Huss, G. R., Alexander, C. M. O'D., Palme, H., Bland, P. A., \& Wasson, J. T. 2005, in Chondrites and Protoplanetary Disk, ed. A. N. Krot, E. R. D. Scott, \& B. Reipurth (San Francisco: Astronomical Society of the Pacific), 701
\bibitem[Iida et al.(2001)]{iida01} Iida, A., Nakamoto, T., Susa, H., \& Nakagawa, Y. 2001, Icarus, 153, 430
\bibitem[Kobatake et al.(2008)]{kobatake08} Kobatake, H., Tsukamoto, K., Nozawa, J., Nagashima, K., Satoh, H., \& Dold, P. 2008, Icarus, 198, 208
\bibitem[Kozasa \& Hasegawa(1987)]{kozasa87} Kozasa, T., \& Hasegawa, H. 1987, Prog. Theor. Phys., 77, 1402
\bibitem[Miura et al.(2002)]{miura02} Miura, H., Nakamoto, T., \& Susa, H. 2002, Icarus, 160, 258
\bibitem[Miura \& Nakamoto(2005)]{miura05} Miura, H., \& Nakamoto, T. 2005, Icarus, 175, 289
\bibitem[Miura \& Nakamoto(2006)]{miura06} Miura, H., \& Nakamoto, T. 2006, ApJ, 651, 1272
\bibitem[Miyake \& Nakagawa(1993)]{miyake93} Miyake, K., \& Nakagawa, Y. 1993, Icarus, 106, 20
\bibitem[Mysen \& Kushiro(1988)]{mysen88} Mysen, B. O., \& Kushiro, I. 1988, American Mineralogist, 73, 1
\bibitem[Nagahara et al.(1996)]{nagahara96} Nagahara, H. \& Ozawa, K. 1996, Geochim. Cosmochim. Acta., 60, 1445
\bibitem[Nagasawa et al.(2005)]{nagasawa05} Nagasawa, M., Lin, D. N. C., \& Thommes, E. 2005, ApJ, 635, 578
\bibitem[Nagashima et al.(2006)]{nagashima06} Nagashima, K., Tsukamoto, K., Satoh, H., Kobatake, H., \& Dold, P. 2006, J. Crys. Growth, 293, 193
\bibitem[Nozawa et al.(2009)]{nozawa09} Nozawa, J., Tsukamoto, K., Kobatake, H., Yamada, J., Satoh, H., Nagashima, K., Miura, H., \& Kimura, Y. 2009, Icarus, 204, 681
\bibitem[Susa et al.(1998)]{susa98} Susa, H., Uehara, H., Nishi, R., \& Yamada, M. 1998, Prog. Theor. Phys., 100, 63
\bibitem[Toriumi(1989)]{toriumi89} Toriumi, M. 1989, Earth Planet. Sci. Lett., 92, 265
\bibitem[Tsuchiyama et al.(1988)]{tsuchiyama88} Tsuchiyama, A., Kushiro, I., Mysen, B. O., \& Morimoto, N. 1988, Proc. NIPR Symp. Antarct. Meteorites, 1, 185
\bibitem[Tsuchiyama et al.(1998)]{tsuchiyama98} Tsuchiyama, A., Takahashi, T., \& Tachibana, S. 1998, Mineralogical Journal, 20, 113
\bibitem[Tsukamoto et al.(1999)]{tsukamoto99} Tsukamoto, K., Satoh, H., Takamura, Y., \& Kuribayashi, K. 1999, Antarct. Meteorites, 24, 179
\bibitem[Wasson \& Rubin(2003)]{wasson03} Wasson, J. T. \& Rubin, A. E. 2003, Geochim. Cosmochim. Acta, 67, 2239
\bibitem[Weidenschilling et al.(1998)]{weidenschilling98} Weidenschilling, S. J., Marzari, F., \& Hood, L. L. 1998, Science, 279, 681
\bibitem[Yamada(2009)]{yamada09} Yamada, J. 2009, Master's Theses, Department of Earth and Planetary Science, Tohoku University
\bibitem[Yamamoto \& Hasegawa(1977)]{yamamoto77} Yamamoto, T., \& Hasegawa, H. 1977, Prog. Theor. Phys., 58,  816
\bibitem[Yamamoto et al.(2001)]{yamamoto01} Yamamoto, T., Chigai, T., Watanabe, S., \& Kozasa, T. 2001, A\&A, 380, 373
\bibitem[Yurimoto \& Wasson(2002)]{yurimoto02} Yurimoto, H. \& Wasson, J. T. 2002, Geochim. Cosmochim. Acta, 66, 4355
\end{thebibliography}
\end{document}